\documentclass[10, journal]{IEEEtran}
\IEEEoverridecommandlockouts

\usepackage{xurl}
\usepackage[hidelinks]{hyperref}
\usepackage{amssymb,amsmath,amsthm}
\usepackage{todonotes}
\usepackage{mathtools}
\usepackage{comment}
\usepackage{dsfont}
\usepackage{ulem}
\usepackage{float}
\usepackage{multirow}
\usepackage{enumitem}
\newtheorem{theorem}{Theorem}

\usepackage{graphicx}
\usepackage{subcaption}
\usepackage{amsthm, amssymb, mathtools, amsmath, xcolor}
\usepackage{booktabs}
\usepackage[noend]{algpseudocode}
\usepackage[ruled]{algorithm}
\algrenewcommand\algorithmicindent{0.8em}

\newcommand{\revise}[1]{\begingroup\color{black}#1\endgroup}

\newcommand{\SC}[1]{\todo[color=yellow,inline]{SC: #1}}

\title{Queuing for Secrets: Model-Based Remote Timing Attacks\thanks{This work has been submitted to the IEEE for possible publication. Copyright may be transferred without notice, after which this version may no longer be accessible.}}
\author{
Simone Bozzolan$^{1}$, Diletta Olliaro$^{2}$, Stefano Calzavara$^1$,\\ Andrea Marin$^1$~\IEEEmembership{Senior Member, ~IEEE}, {Gianfranco Balbo}$^{3}$, {Matteo Sereno}$^3$~\IEEEmembership{Senior Member, ~IEEE}  \\ $^1$ Universita' Ca' Foscari, Venezia, Italy \\ $^2$ KTH Royal Institute of Technology, Stockholm, Sweden \\ $^3$ Universita' degli Studi di Torino, Torino, Italy}

\newcommand{\Ls}{s}

\newcommand{\wcdf}{W}

\newcommand{\wlst}{w^*}
\newcommand{\wrv}{\mathcal{W}}

\newcommand{\zrv}{\mathcal{Z}}

\newcommand{\rcdf}{R}

\newcommand{\rlst}{r^*}
\newcommand{\rrv}{\mathcal{R}}

\newcommand{\xrv}{\mathcal{X}}

\newcommand{\vcdf}{V}
\newcommand{\vpdf}{v}

\newcommand{\vrv}{\mathcal{V}}

\newcommand{\muntzroot}{y}

\begin{document}

\maketitle

\begin{abstract}
Remote timing attacks exploit variations in response times to infer sensitive information about remote systems. Their effectiveness, however, is often limited by the difficulty of accurately modeling the stochastic nature of server-side response times, taking into account the intrinsic variability of the service time of a request and the impact of varying workloads on the observed response times. In this article, we propose a new stochastic framework for remote timing attacks based on queuing models and Bayesian statistics. We assume that the attacker knows the service time distributions of two classes of requests and that the information leakage occurs when a malicious observer can distinguish with high probability the class based on the observed response times in the presence of noise induced by exogenous workload. 

Our experimental evaluation on popular websites demonstrates the practical viability of the approach: our attack outperforms formal techniques based on deterministic service times and existing empirical techniques, while requiring less tuning and fewer hyper-parameters compared to previous formal approaches. Overall, our results show that principled stochastic modeling can provide a solid foundation for building more effective remote timing attacks under realistic network and workload conditions.
\end{abstract}

\begin{IEEEkeywords}
Web Privacy,
Queuing Theory,
Remote Timing Attacks
\end{IEEEkeywords}

\section{Introduction}


Timing attacks exploit timing information to infer private information about a target system. When carried out over the Internet, they are commonly referred to as \emph{remote} timing attacks and have been shown to have significant practical consequences~\cite{CrosbyWR09}. For instance, Brumley and Boneh showed that an attacker could measure response-time variations from a remote web server to reconstruct its RSA (Rivest-Shamir-Adleman~\cite{RSA78}) private key~\cite{BrumleyB05}. In the privacy setting, Felten and Schneider used a cache-based timing attack to track web users~\cite{timingAttacksWebPr}. The underlying observation is that, once a user visits a static page, the browser cache causes the page to load faster on subsequent visits. Later work abused remote timing attacks to disclose sensitive information about target users, such as the existence of active accounts~\cite{BortzB07} and the authentication state of a victim~\cite{bakingTimer}. This line of work is important because it demonstrated the practicality of remote timing attacks, which may otherwise appear far-fetched; however, prior work is largely based on simple empirical observations with limited scientific underpinnings. At a high level, previous attacks rely on the same intuition: attackers collect response times and apply some form of statistical analysis to determine whether they reveal information about the target system. The key differences across proposals lie in the attack objectives, the details of data collection, and the specific nature of the statistical analysis, ranging from simple visual inspection of histograms to the use of statistical tests such as the t-test.

Inspired by the rich research line on queuing theory~\cite{casale18, baccelli2009}, in this article we put forward a novel approach to remote timing attacks based on stochastic modeling. Queuing theory is a popular branch of applied mathematics widely used for performance engineering tasks. It resorts to stochastic models to predict quantitative behaviors of systems such as their response times. At first glance, queuing models may seem to rely on strong and unrealistic assumptions for numerical tractability, but experience in engineering practice has demonstrated their remarkable robustness, allowing practitioners to adopt them for many purposes~\cite{menasce}. A queuing model can predict the expected response time and other quantitative indices of a network request given its type, e.g., a payment request. Here, we propose a novel approach that combines Bayesian statistics with queuing modeling to address the inverse problem: determining the likely type of a network request based on its observed response time.

This approach offers a potential new foundation for remote timing attacks by outlining a general framework that may help distinguish different types of requests based on timing information. Notably, the relevance of this perspective is not purely theoretical; it also offers practical advantages. Our method appears naturally suited to handle the complexities that make performance evaluation challenging, such as the impact of the load factor on the target system’s response times --- factors that can sometimes limit the robustness of purely empirical attacks.

Moreover, our attack establishes quantitative probability estimates about the predicted request type, thus providing a formal assurance about the estimated accuracy of the attack, and leverages the generality of our key idea to be uniformly applied to different settings. Even from a purely experimental perspective, our attack is superior to those proposed in the literature, outperforming the BakingTimer attack~\cite{bakingTimer}.

We conclude that the formal foundations put forward in this article can be successfully applied to the creation of principled remote timing attacks which are more effective in practice, because better equipped to deal with the complexity of the problem they are trying to solve.


\subsection*{Contributions}
In this article, we make the following contributions:
\begin{enumerate}
    \item 
    We propose a new timing attack strategy that builds on the combined use of queuing modeling and Bayesian statistics to improve existing empirical approaches proposed in the literature (Section~\ref{sec:contribution}).
    By leveraging the quantitative analysis supported by the introduced stochastic model, our approach aims to address key limitations in current methods, offering a more robust framework for timing attacks (Section~\ref{sec:attack}).
    \item We propose a specific instantiation of the stochastic model supporting our timing attack based on a M/M/1 Processor Sharing (M/M/1/PS) queuing system~\cite{kleinrock-vol1, mor}. We use standard mathematical tools to support quantitative analyses designed to implement effective remote timing attacks in practice (Section~\ref{sec:model}).
    \item We implement our new timing attack and we compare its effectiveness against those appearing in the state-of-the-art, discussing the key advantages enabled by the adoption of our proposal (Section~\ref{sec:experiments}).
\end{enumerate}

To support reproducibility and fully comply with open science policies, we make the code of our implementation and the collected data available for review~\cite{Artifacts}.

\subsection*{Extensions over Conference Version}

This article substantially reworks the model introduced in our preliminary conference version~\cite{ourpets}. In the conference paper, the model relied on a discretization of the service-time probability distribution. In particular, the distribution was partitioned into $q$ contiguous, non-overlapping intervals (bins), yielding a simplified discrete approximation of the service-time distribution. This approach required the introduction of empirical thresholds to map continuous service-time observations to discrete events, since the probability of observing any exact service-time value is zero in a continuous setting. These thresholds required tuning based on measured response times, adding overhead and reducing robustness due to their sensitivity to measurement variability. In contrast, the model proposed in this article operates directly in continuous time, thereby eliminating the need for discretization and the associated empirical parameters (Section~\ref{sec:model}). This is a significant methodological extension over the previous model: it changes the statistical formulation of the attack, removes an empirically tuned component of the original approach, and enables a more robust treatment of service-time variability. 

This transition to a continuous-time formulation also required addressing several non-trivial methodological challenges. One of the main challenges was identifying a service-time distribution that accurately captures the characteristics of the considered setting while remaining analytically tractable within the proposed inference framework. Furthermore, the continuous formulation led to the development of new mathematical derivations. In particular, the expression derived in Eq.~\ref{eq:rstvariance}, to the best of our knowledge, has not been explicitly derived before in the literature. These developments considerably extend the theoretical foundations of the original work. Moreover, as shown by our experimental evaluation (Section~\ref{sec:experiments}), these methodological advances also translate into performance that is either superior to or on par with the previous approach.

\section{Background}
We here review selected technical ingredients required to appreciate the article. We also introduce the key terminology and notation.

\subsection{Remote Timing Attacks}
In a remote timing attack, the attacker infers information about server-side computations based on measured response times, e.g., to determine sensitive state information about other users of the system~\cite{CrosbyWR09}. More specifically, we assume that each computation belongs to one of $c \geq 2$ possible classes and the attacker measures the response times of network requests to determine the class of the corresponding remote computation. For simplicity, we refer to the class of a request as the class of the associated server-side computation.

To make our investigation concrete, we focus on \emph{cross-site} timing attacks from the web security setting, although our approach is more general and works for any remote timing attack. Cross-site timing attacks circumvent the Same Origin Policy (SOP) of web browsers by letting the owner of a malicious website, e.g., \url{attacker.com}, infer sensitive information about users of a different website, e.g., \url{target.com}; these attacks may be performed by any attacker with the limited capabilities of a traditional \emph{web attacker}~\cite{AkhaweBLMS10}.

In practice, the attack starts when the victim uses their web browser to access \url{attacker.com}. The attack may target casual visitors or abuse social engineering techniques like phishing to lure selected victims. The malicious website uses JavaScript to send requests to \url{target.com} from the victim's browser and measures the response times to infer information about the victim's state on the target server, e.g., using the Performance API. For example, the target web application may implement a private area taking more time to generate a valid profile page than an error page denying access to the profile. This way, the attacker may send a request to the profile page from the victim's browser and measure the response time to infer the victim's authentication state on the target server. These attacks pose a serious threat to user privacy. For example, if an attacker targets a specialized medical center, such as one for HIV treatment or mental health services, determining that a user is logged in could expose highly sensitive health information, as it reveals their active and ongoing medical engagement. This exposure could lead to discrimination, blackmail, or significant distress if the information is misused.

\subsection{Terminology}
This section establishes key concepts from queuing theory, which models systems processing incoming requests.

In a \emph{queuing system}, requests arrive randomly, compete for limited resources, receive service, and then leave the system. A scheduling discipline determines how resources are allocated among requests. The model considered in this article uses the \emph{Processor Sharing} (PS) policy~\cite{kleinrock-vol1, mor}, which evenly distributes computational resources to pending requests. This policy fairly approximates the commonly used Round-Robin scheduling in real systems.

\begin{itemize}[leftmargin=*]
\item \emph{Service time} ($\zrv$) is the time a request would require to be processed if it were the only one in the system. As system internal behaviours (e.g., garbage collection, caching, etc.) introduce variability, service time is treated as a random variable (r.v.). The average number of jobs served per second (i.e., \emph{service rate}) is denoted by $\mu$, thus the average service time is equal to $\mu^{-1}$.

\item \emph{Response time} ($\rrv$) is the total time from when a request arrives in the system until the response is produced.

\item \emph{Waiting time} ($\wrv$) is the time a request spends competing for system resources before receiving service. In a PS system, it represents delays caused by the simultaneous sharing of resources with other requests and it equals the response time minus the service time in the system competing for the server's processing power with other requests, i.e., $\wrv=\rrv-\zrv$.

\item \emph{Arrival rate} ($\lambda$) is the expected number of requests entering the system per unit of time.

\item \emph{Load factor} ($\rho$) measures the system utilization and is defined as the ratio of arrival rate to service rate ($\rho=\lambda / \mu$ for a single-server system).

\item A \emph{stationary system} is one where the load factor remains constant over a sufficiently long period of time. For queuing systems, this typically requires stability (i.e., load factor $\rho < 1$). Although resource usage may fluctuate, the statistical properties of the system remain stable over time. This is a typical assumption in queuing theory as it allows meaningful analysis of waiting and response times. Throughout the rest of this article, we assume that the systems under consideration are stationary.
\end{itemize}

\section{Motivation and Contribution}
\label{sec:contribution}

Remote timing attacks leverage the observation that the response times of network requests may reveal information about server-side computations. Most existing attacks follow the same general strategy: attackers collect response times and apply statistical analysis to determine whether the observed differences reveal information about the target system~\cite{CrosbyWR09,bakingTimer}.

The approach proposed in this article departs from this purely empirical methodology. Instead, it incorporates stochastic modeling, combining queuing theory and Bayesian statistics in a novel way to enhance and complement the information gathered from response times. By doing so, it addresses the limitations of prior work that relied solely on the statistical comparison of raw response-time samples. In particular, the proposed framework accounts for additional system parameters, such as the server load factor, which can significantly influence response times and, consequently, the correctness of the resulting inferences. The proposed approach therefore improves upon the inherent limitations of existing methods, yielding a highly predictive model that enhances the accuracy and practicality of remote timing attacks.


\begin{figure*}[t]
\centering
        \begin{subfigure}[b]{0.35\textwidth}
        \centering
        \includegraphics[width=\textwidth]{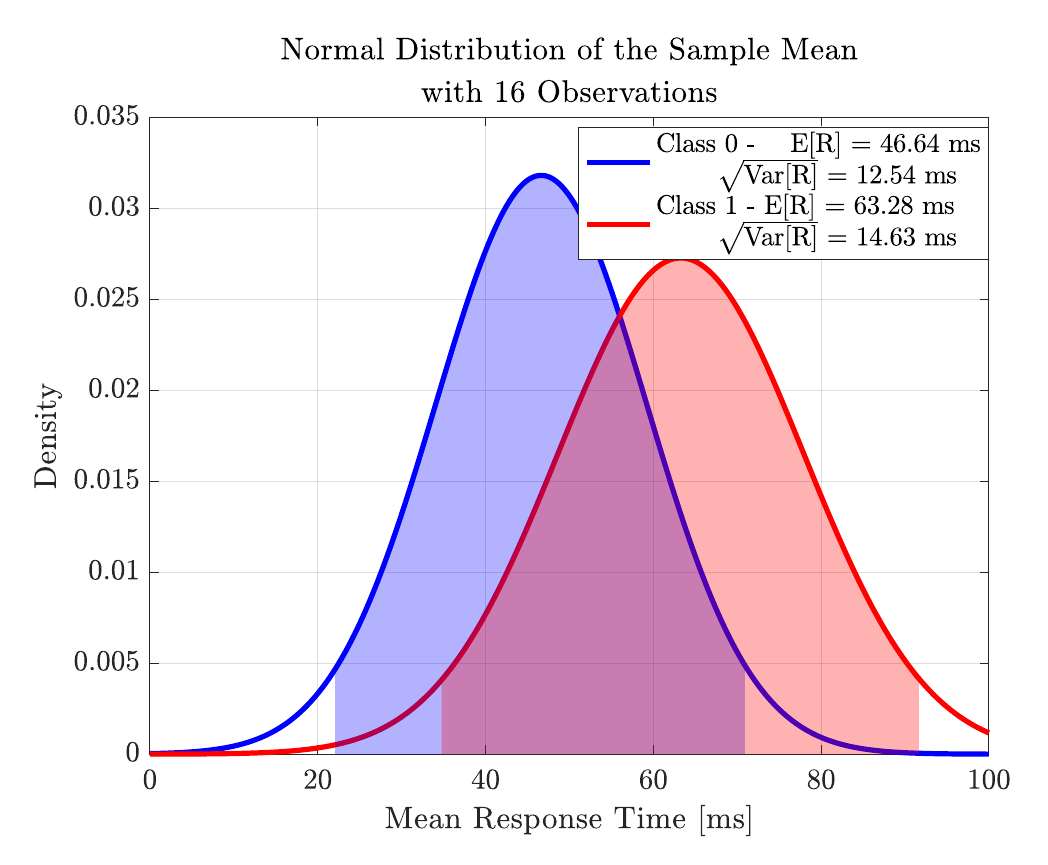}
        \caption{$\boldsymbol{\mu=40}$ j/s, $\boldsymbol{\rho=0.4}$}
        \label{subfig:norm.distro_mu40_rho04_obs4}
    \end{subfigure}
        \begin{subfigure}[b]{0.35\textwidth}
        \centering
        \includegraphics[width=\textwidth]{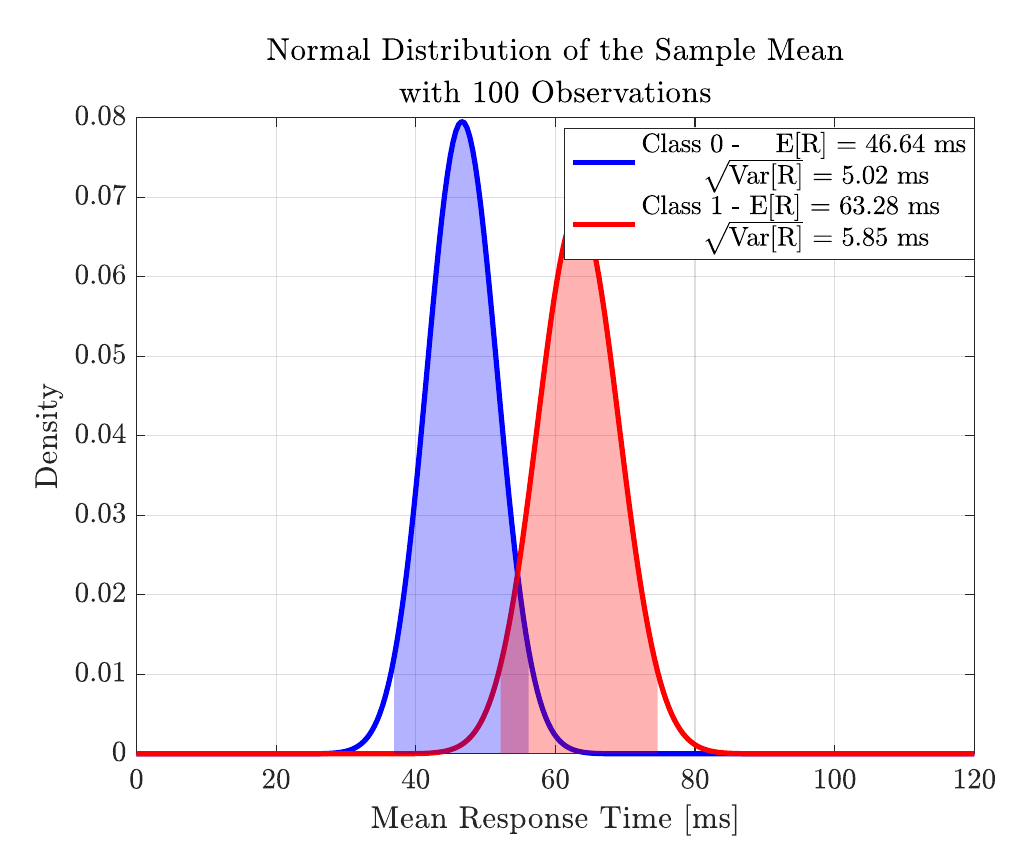}
        \caption{$\boldsymbol{\mu=40}$ j/s, $\boldsymbol{\rho=0.4}$}
        \label{subfig:norm.distro_mu40_rho04_obs10}
    \end{subfigure}

    \begin{subfigure}[b]{0.35\textwidth}
        \centering
        \includegraphics[width=\textwidth]{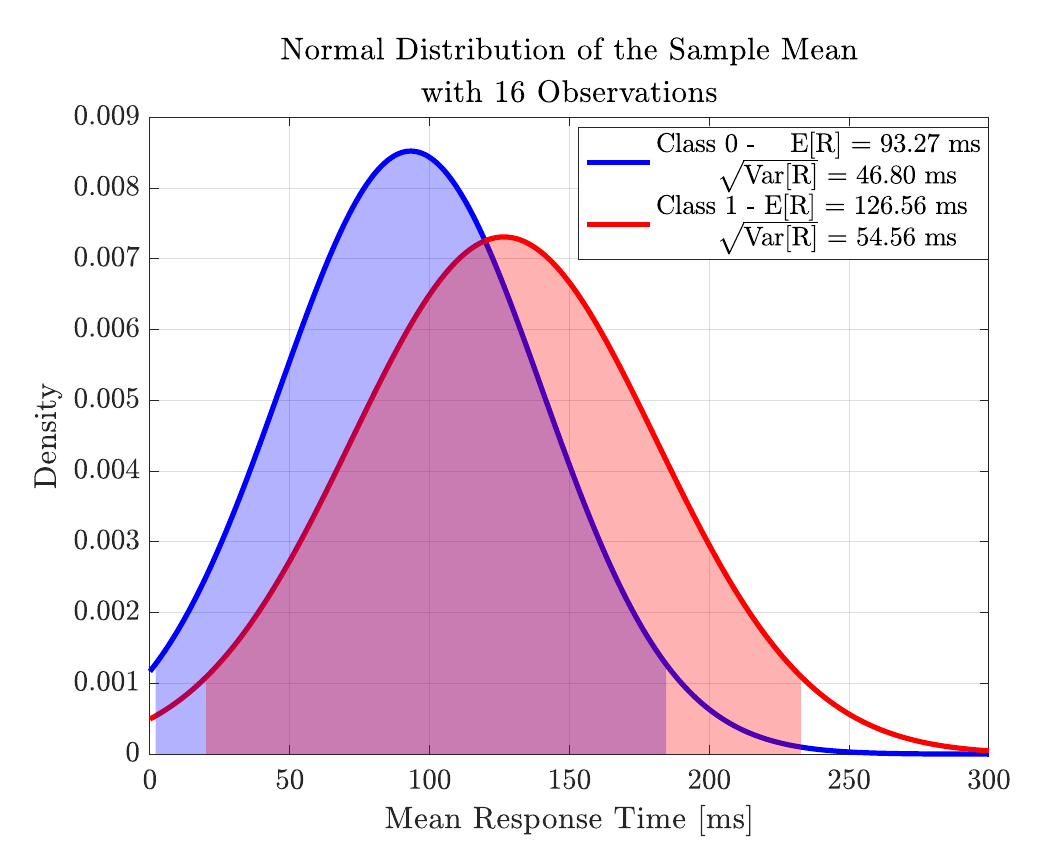}
        \caption{$\boldsymbol{\mu=40}$ j/s, $\boldsymbol{\rho=0.7}$}
        \label{subfig:norm.distro_mu40_rho07_obs4}
    \end{subfigure}
    \begin{subfigure}[b]{0.35\textwidth}
        \centering
        \includegraphics[width=\textwidth]{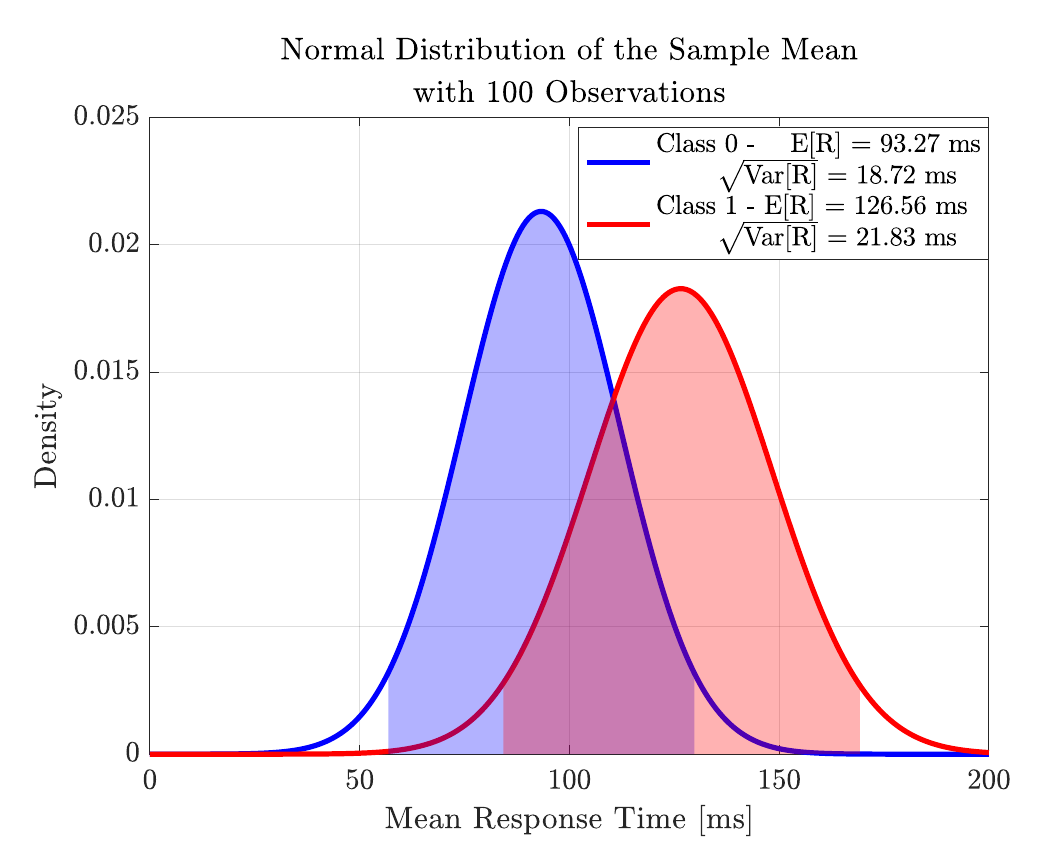}
        \caption{$\boldsymbol{\mu=40}$ j/s, $\boldsymbol{\rho=0.7}$}
        \label{subfig:norm.distro_mu40_rho07_obs10}
    \end{subfigure}

    \caption{Response time distributions under different load factors $\boldsymbol{\rho}$ for a fixed service rate $\boldsymbol{\mu}$.}
    \label{fig:resp.times}
\end{figure*}

\subsection{Response Time vs. Service Time}
A key observation of our work is that timing attacks are difficult to carry out in practice because they abuse differences in the service time, but the attacker can only measure the response time. These two notions roughly coincide when the target web application is under a low load factor ($\rho \approx 0$), but the response time may significantly differ from the service time when the load factor increases and gets closer to 1.
This issue has only been marginally considered in prior work and just from an empirical perspective, e.g., the authors of the BakingTimer attack~\cite{bakingTimer} present a small-scale experiment showing that the time of the day has limited impact on the effectiveness of their attack, but this informal and limited evaluation should be considered with some reservations.

To clarify that our criticism is not just a speculation, Figure~\ref{fig:resp.times} reports the mean response times computed for two classes of requests (authenticated vs. unauthenticated) of a local installation of the popular WordPress web application. Intuitively, the difference in the response times may be abused to carry out a cross-site timing attack breaching the victim's privacy by inferring their authentication state on WordPress. The figure shows that the load factor $\rho$ may have a significant impact on the precision of the attack.

Figures~\ref{subfig:norm.distro_mu40_rho04_obs4} and~\ref{subfig:norm.distro_mu40_rho04_obs10} show the case of a moderate load ($\rho=0.4$), where increasing the number of observations from 16 to 100 improves the ability to distinguish between distributions. On the other hand, Figures~\ref{subfig:norm.distro_mu40_rho07_obs4} and~\ref{subfig:norm.distro_mu40_rho07_obs10} show that under higher load ($\rho=0.7$), the overlap between response time distributions becomes significantly larger, making it difficult to distinguish the classes even with the larger set of 100 observations. This highlights the key limitation of traditional timing attacks: response time measurements alone may become unreliable when server load increases.

\subsection{Proposed Approach}\label{subsec:proposed}

\begin{figure}[t]
    \centering
    \includegraphics[width=0.4\textwidth]{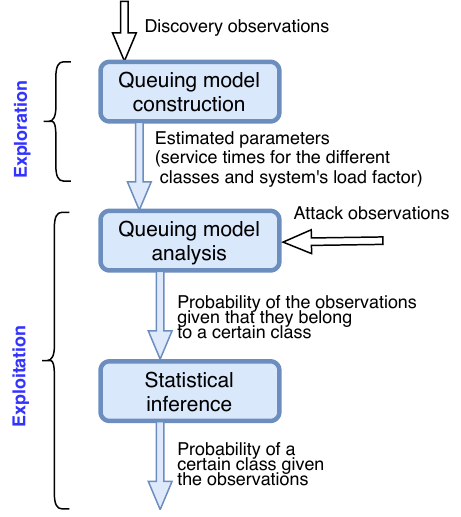}
    \caption{Sketch of the methodology proposed in this article.}\label{fig:methodology}
\end{figure}

Figure~\ref{fig:methodology} sketches the attack methodology proposed in our article. Rather than using the measured response times directly to draw conclusions, the attacker uses them to construct a queuing model of the target web application. Traditionally, queuing systems take the arrival rate $\lambda$ and the service rate $\mu$ as known input parameters to compute key performance metrics such as the system's average response and waiting times. In contrast, with our approach we reverse this process: by collecting response times from a real system, we leverage queuing theory to estimate the unknown parameters $\lambda$ and $\mu$ used to construct an abstract model of the real system. The queuing model obtained from the estimated parameters determines the probability distribution of service times for requests in different classes. This information can then be given as input to a Bayesian statistical model that addresses the inverse question: given a set of observed response times, what is the probability that a request belongs to a particular class?

This approach has many distinctive advantages. To begin with, the model natively incorporates the effects of the system load factor by predicting the response time distribution of network requests based on their service times and the system's load. Consequently, given the response time of a network request observed under a specific load factor, the model can predict the response time distribution of the target network request, whose service time is known, under the same load conditions. It can then estimate the probability that the observed response time belongs to the predicted distribution, enabling accurate probability estimation independently of how the response times are influenced by the load factor. This also makes the attack general, as it applies whenever server-side computations can be partitioned into $c \geq 2$ classes based on their service times, under minimal assumptions about the attacker. The attacker only needs to send HTTP requests belonging to the $c$ classes (e.g., authenticated and unauthenticated) using their own browser, without knowing in advance whether they are actually distinguishable. They then use these observed response times to estimate service times and the load factor, and construct a stochastic model that reveals exploitable differences. Once constructed, the model enables classification of requests based solely on their measured response times.


\section{Attack Description}
\label{sec:attack}
Our attack operates in two phases, called \emph{exploration} and \emph{exploitation}, respectively. During the exploration phase, the attacker collects timing information about the target web application at different times and for different classes of requests. The exploitation phase leverages the information collected in the exploration phase to build a stochastic model of the target web application and actually mount the attack. In particular, the attacker measures the response times of requests of an unknown class and uses the model to reconstruct their class. While we describe the attack assuming that the attacker wants to discriminate between two possible classes for simplicity, the extension to $c > 2$ classes is straightforward.

\subsection{Preliminaries}
We present the two phases of our attack using pseudo-code. Both the exploration phase and the exploitation phase need to measure response times, which are then fed to our stochastic model. However, when we measure the time between an HTTP request and its response, we are also measuring the network round-trip time ($RTT$). Since our model does not account for $RTT$ or its variability, we need to eliminate its impact from our measurements. We achieve this by empirically measuring $RTT$ and subtracting it from our data. To obtain a reliable estimate of response times, we employ two sub-procedures.

The first sub-procedure $\textsc{TimeRequest}(u)$ sends an HTTP request to the URL $u$ and estimates its response time. This sub-procedure is used in the exploitation phase and is implemented as follows. We first send an HTTP request $r$ to $u$ and we measure the time $t_r$ until its response. $t_r$ includes both the response time of $r$ and $RTT$. Along with the request $r$, we craft another one $r'$, whose time until response $t_r'$ serves as a possible estimate of $RTT$. An example of requests that are a suitable choice for this purpose are those unconditionally triggering a redirect. The reason behind this is that such redirects often require negligible server-side computation. One effective way to craft such a redirect request is by modifying the protocol of URL $u$ from HTTPS to HTTP. If the target site does not enforce HTTP Strict Transport Security (HSTS), this HTTP request is likely to trigger a redirect to the HTTPS version of the site. We can then estimate the response time of $r$ as $t_r - t_r'$. Since measurement inaccuracies are inherent to any empirical analysis, subtracting $RTT$ values from the measurements ($t_r - t_r'$) may yield zero or negative response times. To mitigate this issue, we estimate the $RTT$ as the minimum observed value of $t_r'$ across all measurements and subtract it from each measurement. If this adjustment still results in a zero or negative response time, we set the resulting value to $0.1$ to avoid unrealistic results.


The other sub-procedure $\textsc{TimeRequestWithClass}(u,i)$ sends an HTTP request of class $i$ to the URL $u$ and estimates its response time. This sub-procedure is used in the exploration phase, hence it assumes that the attacker is able to craft a request of class $i$. The way the attacker does that depends on the specific class of requests. For example, authenticated requests require the attacker to be able to login to the target web application, while unauthenticated requests can always be forged. The estimate of the response time is performed using the same technique proposed for the $\textsc{TimeRequest}(u)$ sub-procedure, just the format of the request $r$ will be different depending on its class.

Finally, our pseudo-code relies on the construction of a stochastic model $M$, which is queried to determine probability estimates of the class of a target request. Model construction and interrogation are implemented by the sub-procedures $\textsc{BuildModel}$ and $\textsc{EstimateProba}$, which are discussed in Section~\ref{sec:model}. In this section, we consider the model $M$ as a black-box to simplify the presentation.

\begin{algorithm}[ht]
\caption{Exploration Phase}
\label{alg:exploration}
\begin{algorithmic}[1]
\Function{Explore}{$u, endtime, tick, samples$}
\State{$R_0 \gets \{ \}$} \Comment{Response times for class 0 (empty dict.)}
\State{$R_1 \gets \{ \}$} \Comment{Response times for class 1 (empty dict.)}
\State{$b_0 \gets +\infty$} \Comment{Service time estimate for class 0}
\State{$b_1 \gets +\infty$} \Comment{Service time estimate for class 1}
\State{$t_0 \gets 0$}
\State{$t_1 \gets 0$}

\State{$t \gets \Call{GetCurrentTime}{ }$} \Comment{Start time of exploration}
\While{$t \leq endtime$}
\State{$R_0[t] \gets []$} \Comment{$R_0[t]$ is initialized to the empty list}
\State{$R_1[t] \gets []$} \Comment{$R_1[t]$ is initialized to the empty list}

\State{$reqs \gets 0$}
\While{$reqs < samples$}
    \State{$t_r \gets \Call{TimeRequestWithClass}{u, 0}$}
    \State{$R_0[t] \gets t_r :: R_0[t]$} \Comment{Append $t_r$ to the list $R_0[t]$}
    \State{$t_r \gets \Call{TimeRequestWithClass}{u, 1}$}
    \State{$R_1[t] \gets t_r :: R_1[t]$} \Comment{Append $t_r$ to the list $R_1[t]$}
    \State{$reqs \gets reqs + 1$}
\EndWhile
\If{$\Call{Mean}{R_0[t]} < b_0$}
    \State{$b_0 \gets \Call{Mean}{R_0[t]}$}
    \State{$t_0 \gets t$}
\EndIf
\If{$\Call{Mean}{R_1[t]} < b_1$}
    \State{$b_1 \gets \Call{Mean}{R_1[t]}$}
    \State{$t_1 \gets t$}
\EndIf
\State{$t \gets t + tick$}
\State{\Call{WaitUntil}{$t$}} \Comment{Wait until the next tick}
\EndWhile
\State{\Return{$(R_0[t_0],R_1[t_1],R_0[t-tick])$}}
\EndFunction
\end{algorithmic}
\end{algorithm}

\subsection{Exploration Phase}\label{subsec:exploration}
In the exploration phase, the attacker collects timing information about the target web application, which is used to construct the stochastic model underlying the exploitation phase. The exploration phase is done from the attacker's browser. We present exploration before exploitation because model construction is a precondition to exploitation, however the attacker can extend the exploration phase to run it concurrently with the exploitation phase. This way, the attacker constantly collects up-to-date timing information leading to the creation of more accurate models for the exploitation phase.

The $\textsc{Explore}$ procedure shown in Algorithm~\ref{alg:exploration} takes as input the target URL $u$, the end time of the exploration phase $endtime$, a time delta $tick$ defining the frequency of the data collection, and the number of response time estimates $samples$ to collect for each class at each tick. The procedure uses dictionaries $R_0,R_1$ associating to each time tick $t$ a list of response times for the two request classes. It computes $b_0,b_1$, for class 0 and class 1 respectively, by looking for the time ticks $t_0,t_1$ where the mean of the response times is smallest, i.e., the load factor of the target system is the lowest observable. The procedure returns three lists of timing observations: a list of response times for class 0 under the lowest observed load factor for the class ($R_0[t_0]$), a list of response times for class 1 under the lowest observed load factor for the class ($R_1[t_1]$), and a list containing the most recently measured response times for class 0 ($R_0[t-tick]$). The three lists are subsequently passed as inputs $Y^0, Y^1$ and $Y^R$ of the $\textsc{Exploit}$ procedure, presented in the next section.

If the exploration phase starts at time $starttime$, the total number of requests sent to the target URL is $2 \cdot (\lfloor (endtime - starttime) / tick \rfloor + 1) \cdot samples$.
Data collection is performed periodically to be able to reconstruct accurate estimates of the load factor at different times, thereby supporting the construction of models that are always up to date during the exploitation phase.

\begin{algorithm}[t]
\caption{Exploitation Phase}
\label{alg:exploitation}
\begin{algorithmic}[1]
\Function{Exploit}{$u, Y^0, Y^1, Y^R, \alpha, samples$}
\State{$class \gets \bot$} \Comment{The request class is initially unknown}
\State{$Y^A \gets []$} \Comment{Response times used to mount the attack}
\State{$reqs \gets 0$}
\State{$M \gets \Call{BuildModel}{Y^0, Y^1, Y^R}$}
\While{$reqs < samples$}
    \State{$t_r \gets \Call{TimeRequest}{u}$}
    \State{$Y^A \gets t_r :: Y^A$} \Comment{Append $t_r$ to the list $Y^A$}
    \State{$reqs \gets reqs + 1$}
\EndWhile
\State{$(p_0,p_1) \gets \Call{EstimateProba}{M, Y^A}$}
\If{$(p_0 \geq \alpha \vee p_1 \geq \alpha)$}
    \State{$class \gets \arg\max_{i\in \{0,1\}}(p_i)$}
\EndIf
\State{\Return{$class$}}
\EndFunction
\end{algorithmic}
\end{algorithm}

\subsection{Exploitation Phase}
In the exploitation phase, the attacker uses the information collected in the exploration phase to build a stochastic model used to infer the class of the target requests. In cross-site timing attacks, the exploitation occurs from the victim's browser.

The $\textsc{Exploit}$ procedure shown in Algorithm~\ref{alg:exploitation} takes as input the target URL $u$, the lists of timing observations $Y^0$, $Y^1$, and $Y^R$ collected during the exploration phase, a confidence parameter $\alpha \in (0,1]$ that defines the probability threshold required to infer one of the two classes, and the maximum number of requests $samples$ that the attack can perform. The attack begins by using $Y^0$, $Y^1$, and $Y^R$ to build the stochastic model $M$ employed during the attack. Intuitively, $Y^0, Y^1$ are used to estimate the service times of the two classes, while $Y^R$ is used to estimate the most recent load factor. It then collects the timing observations using a loop that terminates once the maximum number of requests has been sent. At each iteration, the response time at $u$ is estimated and appended to the list of timing observations $Y^A$. This list is later used by the stochastic model $M$ to compute the probabilities $(p_0, p_1)$ of the two classes. If either probability reaches the confidence threshold $\alpha$, the corresponding class is returned. Otherwise, if the desired confidence is not reached, the unknown class $\bot$ is returned.


\section{Stochastic Models}\label{sec:model}
We finally present the formal models used in our attack. We first introduce the queuing model used to represent the target web application. Since this model needs to be initialized with appropriate parameters to predict response times, we then discuss how we estimate such parameters from black-box observations of the target web application. Finally, we present the statistical model that leverages the queuing model to answer the inverse question, i.e., what is the likely class of a request, given an estimate of its service time? These two models are those used by the $\textsc{BuildModel}$ and $\textsc{EstimateProba}$ sub-procedures, respectively (Section~\ref{sec:attack}).

\subsection{Queuing Model}
We model the system under consideration with an M/M/1 queuing model~\cite{kleinrock-vol1,mor}. This means that we assume that the system is composed of one single server. Although web applications are often distributed and run on several servers, simple queuing models can successfully capture the quantitative behavior of these systems by assuming that the single-server abstraction accounts for the entire processing power of the real system~\cite{cao,ConnieW01}. Indeed, this simplification remains highly effective for many real-world websites, as also confirmed by our approach and evaluation.


Arriving jobs follow a Poisson process, with arrival rate $\lambda$, and are served according to a Processor Sharing (PS) policy. Their service times are instances of a r.v. characterized by a negative exponential distribution with mean $E[\zrv] \,(=\mu^{-1})$, so that $\mu$ is the jobs' service rate. Both the characteristics of the system being subject to a Poisson arrival process and the service time being negative exponentially distributed are standard assumptions that allow tractability and have been shown to be robust~\cite{Henry11,ConnieW01}.

The M/M/1/PS queuing system has been extensively studied in the literature, and as a result, many key performance metrics have well-known analytical expressions, under the assumption that the system is stationary. This assumption is not only standard in the queuing theory literature, but also reasonable in this particular setup, because load variations tend to fluctuate over time, but do so quite slowly with respect to service times~\cite{KistowskiHZKH15}.



\paragraph{The Observability Challenge} Once we decided the representative model for the target system, we had to address a fundamental challenge: to be able to use the model itself, we need to be able to parametrise it. This means we must determine the model's arrival rate $\lambda$, the service rate $\mu$, and thus the resulting system load $\rho$. However, in a realistic attack scenario such as the one we are trying to address, we operate under some strict black-box constraints: (i) we cannot directly observe the system arrival rate or its processing capacity, (ii) we are not able to measure the service times of the requests, but only their total end-to-end response times.

To overcome this challenge, we introduce a {\em probe} job into the system. Specifically, we assume that within the stream of statistically identical jobs that represents the input of our M/M/1/PS system, we inject a job with known service characteristics, different from those of the other jobs, and whose response time distribution becomes the objective of our analysis. \\

To ensure that the probe jobs do not disrupt or alter the system states we wish to measure, we inject them at an extremely low frequency. This guarantees the overall impact of the probe jobs on the system's statistics is negligible.
Under this non-intrusive assumption, the background system behavior remains practically unaffected, allowing us to safely apply standard stationary queuing equations to characterize the probe’s performance\footnote{These approximation assumptions have been successfully validated setting up a simulation experiment in which the performance figures of the probe jobs are directly measured within a very long sequence of basic requests from a large number of probe jobs processed by the M/M/1/PS model}.

Specifically, under the assumption that the system is stationary and that $\rho = \dfrac{\lambda}{\mu} < 1$,  the expected value of the response time of a job whose service time is equal to the deterministic quantity $\eta$ is:
\begin{equation}\label{eq:mm1ps-respTime}
    E[\rrv|\eta] = \dfrac{\eta}{1-\rho}\,.
\end{equation}

As we mentioned in Section \ref{subsec:proposed}, we reverse the use of the M/M/1/PS model in this context inverting the equation above to obtain:
\begin{equation}\label{eq:rho}
   \rho = 1 -  \dfrac{\eta}{E[\rrv|\eta]}\,,
\end{equation}

This equation says that once we are able to identify a job of size $\eta$ and if we are able to obtain the value of $E[\rrv|\eta]$, we can compute the system load factor $\rho.$

\paragraph{From Deterministic to Stochastic Service Times} In practical scenarios, assuming a deterministic service time $\eta$ is unrealistic because real-world execution times are subject to several system factors (e.g., resource contention, context-switch overhead, etc.). Thus, to ensure the robustness of our methodology, we model the probe's service time as a continuous random variable $\vrv$ governed by a Gamma distribution, with $\vcdf$ denoting its cumulative distribution function (CDF), and with $\vpdf$ its density function.

The Gamma distribution is defined strictly over non-negative values (as required for service times), and its flexibility (via its shape and rate parameters) allows one to accurately capture the variability observed in the service times of the probe job requests while remaining analytically tractable.

As we shifted our perspective from a deterministic service time to the random variable $\vrv$, we can no longer use the deterministic equation~\eqref{eq:rho} to estimate $\rho$ directly. Instead, we must decondition the expected response time in Eq.~\eqref{eq:mm1ps-respTime} over the entire distribution of the probe's service time in the following manner:

\[
E[\rrv] = \int_0^{\infty} E[\rrv|\eta] \, \vpdf(\eta)\ d\eta = \dfrac{1}{1-\rho}\,E[\vrv]\,.
\]
from which, we can easily derive the expression of $\rho$ that we are interested in:
\begin{equation}\label{eq:rhop}
   \rho = 1 -  \dfrac{E[\vrv]}{E[\rrv]}\,.
\end{equation}

While Eq.~(\ref{eq:rhop}) provides a robust estimate of the load factor $\rho$ using quantities that can be measured directly (as discussed in Section~\ref{subsec:params-estimation}), it is insufficient to fully parameterize our model. Since $\rho$ only determines the ratio between the arrival and service rates, an additional independent equation is thus required to identify $\lambda$ or $\mu$ individually. Moreover, as shown in Section~\ref{subsec:statmodel}, our inference framework relies on the full CDF derived from the PDF 
of the probe job's response time rather than on its mean only. 

To address both requirements above, we analyze the response time of the Gamma-distributed probe in the transform domain. In the next subsection, we derive its Laplace-Stieltjes Transform (LST), $\rlst(\Ls)$, which plays two key roles: (i) differentiating the LST yields a closed-form expression for the response-time variance, providing the additional equation required to retrieve $\lambda$ (and consequently $\mu$), (ii) we numerically invert the LST to obtain the exact conditional response-time PDFs used by our Bayesian statistical classifier.

\subsection{Response time distribution and its moments}

Consider a standard M/M/1/PS queuing model and let 
$\mathcal W_\eta$ be the r.v. denoting the (stationary) waiting time of a job with service time equal to $\eta$.
From \cite[Corollary 1]{muntz-lstPSwaitTime} we know that when $\rho<1$, the CDF of this r.v. can be expressed as follows:
\[
\wcdf(\eta, t)=Pr\{\mathcal \wrv_\eta \leq t\}\,.
\]

Then, denoting with $w(\eta,t)$ the corresponding density function, we can define the Laplace-Stieltjes transform:

\[
\wlst(\eta,\Ls)\triangleq\int_0^\infty \text{e}^{-\Ls t} w(\eta,t)\, dt
\]

of the equilibrium waiting time distribution conditioned on the job requiring $\eta$ units of service, namely

\begin{equation}\label{eq:conditional.ps}
\wlst(\eta, \Ls) = \dfrac{(1-\rho)(1-\rho \muntzroot(\Ls)^2)\, \text{e}^{- a(\Ls) \eta}}{b(\Ls) - g(\Ls) \, \text{e}^{- h(\Ls) \eta}}\,.
\end{equation}

where
$\muntzroot(\Ls)$ is the smallest of the two positive roots of the equation
\[
\rho \muntzroot(\Ls)^2 - (\lambda + \mu + \Ls) \muntzroot(\Ls) + \mu = 0
\]
and where the following expressions are introduced to simplify notation:
\begin{align*}
    a(\Ls) &= \lambda(1-\muntzroot(\Ls))\,,\  \quad
    b(\Ls) = (1-\rho \muntzroot(\Ls))^2\,,\\
    g(\Ls) &= \rho (1-\muntzroot(\Ls))^2\,, \quad  h(\Ls) = \dfrac{\mu (1-\rho \muntzroot(\Ls)^2)}{\muntzroot(\Ls)}\,.
\end{align*}

The expression for $\muntzroot(\Ls)$ is the following
\[
\muntzroot(\Ls) = \dfrac{\lambda + \mu + \Ls - [(\lambda+\mu+\Ls)^2 - 4 \mu\lambda]^{1/2}}{2\lambda}\,.
\]

Let $\rrv_\eta$ be the r.v. denoting the response time of a job of size $\eta$, and let $\rcdf(\eta,t)$ be the corresponding CDF defined as:

\[
\rcdf(\eta,t)=
\begin{cases}
0 & \text{if }t < \eta \\
\wcdf(\eta, t-\eta) & t \geq \eta
\end{cases}
\]
i.e., the CDFs of $\rrv_\eta$ and $\wrv_\eta$ are the same modulo a shifting on the $t$-axis of width $\eta$.
\[
\wcdf(\eta, t) = \rcdf(\eta, t+\eta)\,, \quad t\geq \eta
\]

Applying the \textit{time-shifting} property of Laplace transforms, 
we obtain the Laplace transform of the response time, conditioned on the job requiring a deterministic service time $\eta$:
\begin{equation}\label{eq:shift}
\rlst(\eta, \Ls) = \text{e}^{-\Ls \eta} \wlst(\eta, \Ls)\,.
\end{equation}

In order to decondition $\rlst(\eta, \Ls)$ and obtain $\rlst(\Ls)$, which takes into account the distribution of the service time, we follow the reasoning of~\cite{morrison}. Here, an explicit expression for $\rlst(\Ls)$ is obtained for the simpler case of an arbitrary job of the input stream with negative exponential distribution. Denoting with $z(\eta)$ the probe job's service time density function, the explicit expression derives from deconditioning the expression above.

By the law of total probabilities we obtain:
\begin{equation}\label{eq:totprob}
    \rlst(\Ls) = \int_0^\infty \rlst(\eta, \Ls)\, \vpdf(\eta)\, d\eta\,.
\end{equation}

Assuming instead that the job of size $\eta$ is a probe job with known distribution, different, in general, from that of the jobs that are part of the input Poisson stream,  we obtain a result that is useful for our application.


\begin{theorem}\label{th:probe-gamma}
    Given an M/M/1/PS queuing system with $\rho < 1$, assume that a probe job has a service time distributed according to a Gamma r.v. with shape $k$ and rate $k\gamma$. Then, the Laplace-Stieltjes transform of the probe job's response time is:
    \begin{align}\label{eq:ps.response.transform.gamma}
        \rlst(\Ls) &= (k\gamma)^k(1-\rho)(1-\rho \muntzroot(\Ls)^2) \nonumber \\
        &\cdot \dfrac{1}{b(\Ls)h(\Ls)^k}\Phi\left(\dfrac{g(\Ls)}{b(\Ls)}, k, \dfrac{\Ls+a(\Ls)+k\gamma}{h(\Ls)}\right)\,,
    \end{align}
where $\Phi$ is the Lerch transcendent function.

\end{theorem}

\begin{proof}[Sketch of Proof]
The result is established by conditioning on the Gamma-distributed service time of the probe job and integrating its conditional response time transform. The resulting integral is then evaluated by transforming it into a series representation that directly yields the final expression.
The complete algebraic derivations and supporting lemmas are deferred to the supplementary material.
\end{proof}

From the LST of the response time of the probe job, it is possible to derive its analytical variance. Its algebraic derivation is omitted here for brevity, and the complete computations are provided in the supplementary material. It is given by:

\begin{multline}\label{eq:rstvariance}
    \text{Var}({\mathcal R}) = 
    \left.\frac{d^2 \rlst(\Ls)}{d \Ls^2}\right|_{s=0}-\left(\left.\frac{d \rlst(\Ls)}{d \Ls}\right|_{s=0}\right)^2 \\
    =
    \dfrac{1}{\gamma^2 k \left(\lambda - \dfrac{\lambda}{\rho}\right)^4} \left(\dfrac{\lambda}{\rho}\right)\bigg(\left(\lambda - \dfrac{\lambda}{\rho}\right)^2\left(\dfrac{\lambda}{\rho}\right)\\
    + 2\gamma k \lambda 
     \left(\dfrac{\lambda}{\rho} - \lambda\right)+ 2\gamma^2 k \lambda\left(\dfrac{\gamma k}{\gamma k - \lambda + \dfrac{\lambda}{\rho}}-2\gamma^2 k \lambda\right)^k\bigg)\,.
\end{multline}

The importance of this result is that the variance of the response time of the probe job depends, as we mentioned before, solely on the values of $\rho$, $\lambda$, and from the (known) parameters of the distribution of the probe job service time, so that an expression for $\lambda$ can be obtained by inverting this result.


It is worth noting that, although the derivations leading to Eqs.~\ref{eq:ps.response.transform.gamma} and~\eqref{eq:rstvariance} are entirely mechanical rather than conceptual, to the best of our knowledge the resulting explicit expressions have not previously appeared in the literature.

\subsection{Parameter Estimation}\label{subsec:params-estimation}
In this section, we present the methodology that we use to configure our queuing model. In particular, we need to estimate both the system parameters ($\lambda$ and $\mu$) and the service time distribution parameters for the two request classes (class $0$ and class $1$) we want to be able to distinguish between. This is achieved by collecting three specific sets of measurements:

\begin{itemize}
    \item $Y^{0} = (Y_1^{0}, Y_2^{0}, \dots, Y_{N^{0}}^{0})$: response times measured for requests of class $0$ under the lowest observed load factor.
    \item $Y^{1} = (Y_1^{1}, Y_2^{1}, \dots, Y_{N^{1}}^{1})$: response times measured for requests of class $1$ under the lowest observed load factor.
    \item $Y^{R} = (Y_1^{R}, Y_2^{R}, \dots, Y_{N^R}^{R})$: response times for requests of class 0 under the most recent load factor.
\end{itemize}

Since, as previously mentioned, the system appears as a black-box to the attacker, it is not possible to precisely determine the system's load factor at any given moment or other quantitative characteristics. Consequently, parameter estimation relies on a series of reasonable assumptions and leverages properties of the collected observations.

As discussed in Section~\ref{subsec:exploration}, the vectors $Y^{0}$ and $Y^{1}$ are inferred from a subset of observations with the lowest averages. These observations are selected to represent conditions of zero waiting time, providing the most accurate estimates of the service times for the classes under minimal load conditions. Similarly, the vector $Y^R$ is taken from the most recent set of observations, capturing the system's latest operational state. This approach facilitates the estimation of service times for both classes and the system's load factor using the methodology outlined below.

\paragraph{Estimation of the service time distributions} To model the service time of each class, we fit a Gamma distribution to our low-load measurement vectors. Since the estimation methodology is identical for both classes, we drop the class superscripts ($0$ and $1$) and present the derivation for a generic low-load observation vector $Y$. 

Let $\vrv$ denote the service time of the requests, modeled as a Gamma random variable with shape parameter $k$ and mean $1/\gamma$ (thus, rate $k\gamma$):
\begin{equation}
    \vcdf(t, k, \gamma) = \text{Pr}\{\vrv\leq t\} = 1 - \sum_{n=0}^{k-1} \dfrac{(k\gamma t)^n}{n!} e^{-k\gamma t}\,.
\end{equation}

In particular, assuming negligible queuing delay under low-load conditions, the observed response times in $Y$ directly reflect the service times. Consequently, we estimate the parameters $k$ and $\gamma$ by applying Maximum Likelihood Estimation (MLE) to fit a Gamma distribution directly onto the samples collected in $Y$.

\paragraph{System parameters estimation} Once we characterized the service time distributions, we can now estimate the system load factor, $\hat{\rho}$, 
using the measurements contained in $Y^R$. 

From Eq.~(\ref{eq:rhop}), remembering that $E[\vrv]=1/\gamma$, 
and by equating the sample mean of the observations to the analytical mean,
$E[\rrv]=\overline{Y^R}$, we can obtain $\hat{\rho}$ as follows:
\[
\widehat{\rho} = 1 - \dfrac{1}{\gamma\overline{Y^R}}\,.
\]

Following a similar reasoning, in order to estimate the arrival rate, we equate the empirical variance computed from the observations $Y^R$, i.e., $\text{Var}(Y^R)$, to the analytical variance $\text{Var}({\mathcal R})$ given in Eq.~(\ref{eq:rstvariance}). 

Because Eq.~(\ref{eq:rstvariance}) contains transcendental terms, an explicit closed-form expression for $\lambda$ cannot be analytically retrieved. Instead, we employ a numerical root-finding procedure provided by MATLAB to find the zeros of the objective function $\text{Var}({\mathcal R}) - \text{Var}(Y^R)$ and retrieve $\widehat{\lambda}$. Once $\widehat{\lambda}$ is determined, it is straightforward to compute $\widehat{\mu}$ from the load factor relationship given by $\rho=\lambda/\mu$.
The methodology described in this section is included in Algorithm~\ref{alg:exploitation} through the $\textsc{BuildModel}$ sub-procedure.
\subsection{Statistical Model}\label{subsec:statmodel}
In this section, we present the statistical model designed to estimate the posterior probability that an observed set of response times belongs to a specific request class, based on observations of response times and system parameters estimated in the previous section. This model will be used by the $\textsc{EstimateProba}$ sub-procedure (Algorithm~\ref{alg:exploitation}) which implements the evaluation of Eq.~(\ref{eq:posterior.cont}), i.e., the Bayesian classification framework. Hereafter, we show the mathematical derivation of the result.

We let $\Theta \in \{0,1\}$ be the (hidden) class index indicating the active service-time distribution. For each class $i \in \{0,1\}$, we parameterize the Laplace-Stieltjes transform (LST) of the response-time distribution, $\rlst_i(\Ls)$, using the background queuing parameters $\widehat{\lambda}$ and $\widehat{\mu}$ estimated in Section~\ref{subsec:params-estimation}, along with the class-specific service time distribution parameters $(\gamma_i, k_i)$.


Theoretically, the conditional probability density function (PDF) of a single observed response time $y$ can be obtained by taking the inverse Laplace transform of the response time LST evaluated exactly in $y$:
\begin{equation}\label{eq:bayes.pdf.cont}
f_{Y}(y\mid \Theta=i) = \mathcal L^{-1}\!\big\{\rlst_i(s) \big\}\!(y)\,,
\end{equation}
where $\mathcal L^{-1}$ denotes the inverse Laplace transform.

However, evaluating Eq.~\eqref{eq:bayes.pdf.cont} directly presents severe numerical challenges. This is because in an M/M/1/PS system, there is a non-zero probability that an arriving job experiences zero waiting time, meaning the response time is exactly equal to the service time. Consequently, the theoretical PDF $f_Y(y \mid \Theta=i)$ contains a sharp mathematical discontinuity at the minimum service time boundary. Directly inverting the LST of the PDF near this boundary would trigger severe numerical oscillations.

To resolve both challenges, we shift our numerical inversion from the PDF to the CDF, $F_R(y \mid \Theta=i)$, which is continuous, bounded, and numerically stable to invert. We then evaluate the practical PDF using a central-difference numerical derivative smoothed over a small positive interval $\epsilon$:
\begin{equation}\label{eq:bayes.pdf}
f_{Y}(y \mid \Theta=i) \approx \dfrac{F_R(y + \epsilon \mid \Theta=i) - F_R(y - \epsilon \mid \Theta=i)}{2\epsilon}\,,
\end{equation}
where the CDF is computed via numerical inversion:
\begin{equation}\label{eq:cdf.inversion}
F_R(y \mid \Theta=i) = \mathcal L^{-1}\!\left\{\dfrac{1}{\Ls}\rlst_i(\Ls) \right\}\!(y)\,.
\end{equation}
Basically, Eq.~(\ref{eq:bayes.pdf}) is a smoothed PDF with respect to
$\partial F_{R}(x|\Theta)/\partial x$ that does not change the mean of the r.v. $R$ given $\Theta$. 

To compute the inverse Laplace transform in Eq.~\eqref{eq:cdf.inversion}, we employ the Concentrated Matrix Exponential (CME) method~\cite{horvath20}. This technique is known for its exceptional numerical stability and smooth approximation compared to other widely used techniques. We leverage the robust MATLAB package provided in~\cite{matlab_ilt} to compute the joint likelihood of a vector of $N^A$ observed response times $Y^A = (y_1^A, \dots, y_{N^A}^A)$ belonging to class $i$:
\begin{equation}\label{eq:bayes.likelihood.cont}
g_{Y^A}(\Theta=i\mid y_1^A,\dots, y_{N^A}^A) = \prod_{n=1}^{N^A} f_Y(y_n^A\mid \Theta=i)\,.
\end{equation}

Letting $\pi_i \equiv \Pr(\Theta=i)$ denote the prior probability of observing class $i$, i.e., of observing $\Theta = i$, we apply Bayes' theorem to compute the posterior probability that $\Theta = i$ given the observations $Y^A$, i.e., the observations collected during the exploitation phase:
\begin{equation}\label{eq:posterior.cont}
\Pr(\Theta=i\mid y^A_{1}, \dots,  y^A_{N^A}) = \dfrac{g_{Y^A}(\Theta=i\mid y_1^A, \dots, y_{N^A}^A)\pi_i}{G}\,,
\end{equation}
where the normalization denominator is:
\[
G = \sum_{i\in\{0, 1\}} g_{Y^A}(\Theta=i\mid y_1^A, \dots, y_{N^A}^A)\pi_i
\]
This posterior probability serves as the classification metric in our attack.

\subsection{Model Implementation Details}

An important aspect of the implementation of the presented model is ensuring physically meaningful outputs, particularly in scenarios where certain conditions cannot occur. For instance, since waiting times cannot be negative, any negative value due to measurement problems that may be given as input to the numerical inversion is automatically set to zero, reflecting the fact that negative waiting times have zero probability of occurrence. Similarly, if measurement errors lead to an estimated load factor of zero, we bypass the numerical transform inversion entirely. Instead, we use the cumulative density function to deal with the observed timings as service times of one or the other class. Additionally, to prevent probabilities produced by the inverse Laplace transform from assuming zero or near-zero values, which otherwise would allow a single response time to dominate or disproportionately affect the final result despite the contribution of all other observations, we constrain the output probabilities to the interval $[0.02, 0.98]$. 
Finally, to account for measurement uncertainty, which is inherent to any empirical analysis, we introduce a threshold $\epsilon$ (used in Eq.~(\ref{eq:bayes.pdf})) that defines an interval within which the true response time is expected to fall. In our experiments, we set this threshold to 15, since response times are on the order of hundreds of milliseconds, consistent with the value used in our original work~\cite{ourpets}.

\section{Experimental Evaluation}
\label{sec:experiments}

\revise{In this section, we demonstrate our attack in the wild and compare its effectiveness against state-of-the-art baselines. From an ethical perspective, because our attack is tested on real-world websites, we took care to minimize potential harm and ensure that the benefits outweigh the risks. In particular, all our attacks only targeted our own accounts, without breaching the privacy of other web users. Furthermore, all identified privacy issues were responsibly disclosed to the appropriate contacts for their consideration.}


\subsection{Experimental Setup}


\revise{We evaluate the effectiveness of our attack on 20 popular websites from the Tranco ranking~\cite{PochatGTKJ19}, assessing its performance in real-world settings and its ability to successfully target previously unseen web applications deployed on the Internet. Our evaluation considers cross-site timing attacks, where the attacker's goal is to disclose the class of an unknown request. We consider just two request classes, which we refer to as positive (authenticated) and negative (unauthenticated).}


\subsubsection{Baselines}
\revise{As a first baseline to compare against, we re-implemented BakingTimer~\cite{bakingTimer}, a state-of-the-art cross-site timing attack. BakingTimer is a specific instantiation of a general attack strategy in which the adversary collects sets of timing observations and applies statistical tests to infer the request class. More concretely, the attacker first collects sets of timing observations $U_1,U_2$ for the two classes to discriminate. If a statistical test determines that $U_1$ and $U_2$ likely come from different distributions, the attack can take place. The attacker then collects a set of timing observations $D$ of some unknown class, which they want to disclose. Finally, $D$ undergoes a statistical test against both $U_1$ and $U_2$. If exactly one of the two tests succeeds, the attacker performs the prediction recommended by the successful test, otherwise there is not enough statistical evidence to predict any class (abstention).

In the specific case of BakingTimer, the set $D$ contains ``time deltas'' estimating the server-side computation time in terms of the differences between a set of response times and a set of estimates of the network round-trip time, similar to what we do in our attack. Computing the set of time deltas $D$ requires sending $2 \cdot |D|$ requests. BakingTimer uses the classic t-test to disclose the class of network requests in a cross-site setting, based on the time deltas $U_1,U_2$ of the two target classes.}

As a second baseline, we compare against a version of our attack based on the stochastic model presented in our conference paper~\cite{ourpets}, which, unlike the new model, relies on a discretization of the service time probability distribution. Specifically, starting from the array of observed response times used to estimate the service time for each class, we consider the range between the minimum and maximum observed values and partition it into $q$ contiguous, non-overlapping intervals, namely bins. For each bin, we count the number of observations it contains to estimate how likely that range is, and we represent each bin using its centroid together with the corresponding probability. This yields a simplified, discrete approximation of the service time distribution for each class. The resulting centroids and probabilities are then used to estimate system parameters and, in turn, to compute the probability that a series of response times belongs to a given class. 


Additionally, in the original model we have a threshold $\delta$ to account for the fact that the exact service time centroid is unlikely to match the observed value exactly, due to measurement errors. This threshold defines a small interval within which observations are considered comparable to those associated with zero waiting time. Aside from these two additional hyper-parameters, the interface of the new and old model is the same, so we use the same arrays of observed response times collected using the procedures described in Section~\ref{sec:attack} for both models.


\subsubsection{Hyper-Parameters}
\revise{For BakingTimer, we set the significance level for the t-test at 0.05, as it is a standard recommended practice, and we compare the resulting p-value to this threshold to determine statistical significance. The number of requests is set to 20 because the original paper reports satisfactory performance for any number between 10 and 50~\cite{bakingTimer}. We chose 20 requests as it represents a compromise between attack effectiveness and realism, considering the victim may remain on the attacker's page just for a short duration. Note that the use of 20 requests in the attack actually requires sending 40 requests, organized into 20 pairs, to compute the time deltas.}

For our attack, we set the probability threshold $\alpha$ to 0.90, i.e., we require that the predicted class is assigned a probability of at least 90\%, and the maximum number of HTTP requests to 20, the same number as BakingTimer to enable a fair comparison. The additional parameters required by the original model are set to the same values as in the conference paper~\cite{ourpets}, namely the threshold $\delta$ is set to 0.05 and the number of bins is set to 4.

\subsection{Experiments in the Wild}
\label{subsec:exp-wild}
In cross-site timing attacks, the exploration is carried out from the attacker's browser, while the exploitation occurs from the victim’s browser. To conduct these experiments, we used two separate computers, one for the victim and one for the attacker, both connected to the same wired network and sending requests to the target website over the Internet. The attacker's goal is to determine whether a target user is logged in to the web application or not by collecting response times. 

We compare the effectiveness of our attack against BakingTimer~\cite{bakingTimer} and a variant of our attack based on the original stochastic model~\cite{ourpets}. We focus our investigation on the top 20 websites of the Tranco ranking~\cite{PochatGTKJ19} providing access to a private area and we assess their vulnerability to cross-site timing attacks. We manually navigate websites in decreasing order of popularity until we identify a login form, discarding websites where a login form is not identified within a reasonable amount of time or where registration cannot be readily performed, e.g., when the website requires a paid subscription. For each website, we register a new account (if we do not already have one) and we perform a login. After accessing the private area, we delete from the browser all the cookies whose SameSite attribute is not set to None and we manually check whether we are still logged in to the website. Websites that fail this manual test are not vulnerable to cross-site timing attacks, because their sessions cannot be established through cross-site requests~\cite{KhodayariP22}. Perhaps surprisingly, 13 out of 20 analyzed websites establish sessions using cookies with the SameSite attribute deactivated. \revise{We compare the performance of the different attack methods by conducting 16 attack attempts for each potentially vulnerable website (eight in the authenticated state and eight in the unauthenticated state), leading to 208 attack attempts.}

\begin{figure*}[t]
    \centering  
    \includegraphics[width=0.9\textwidth]{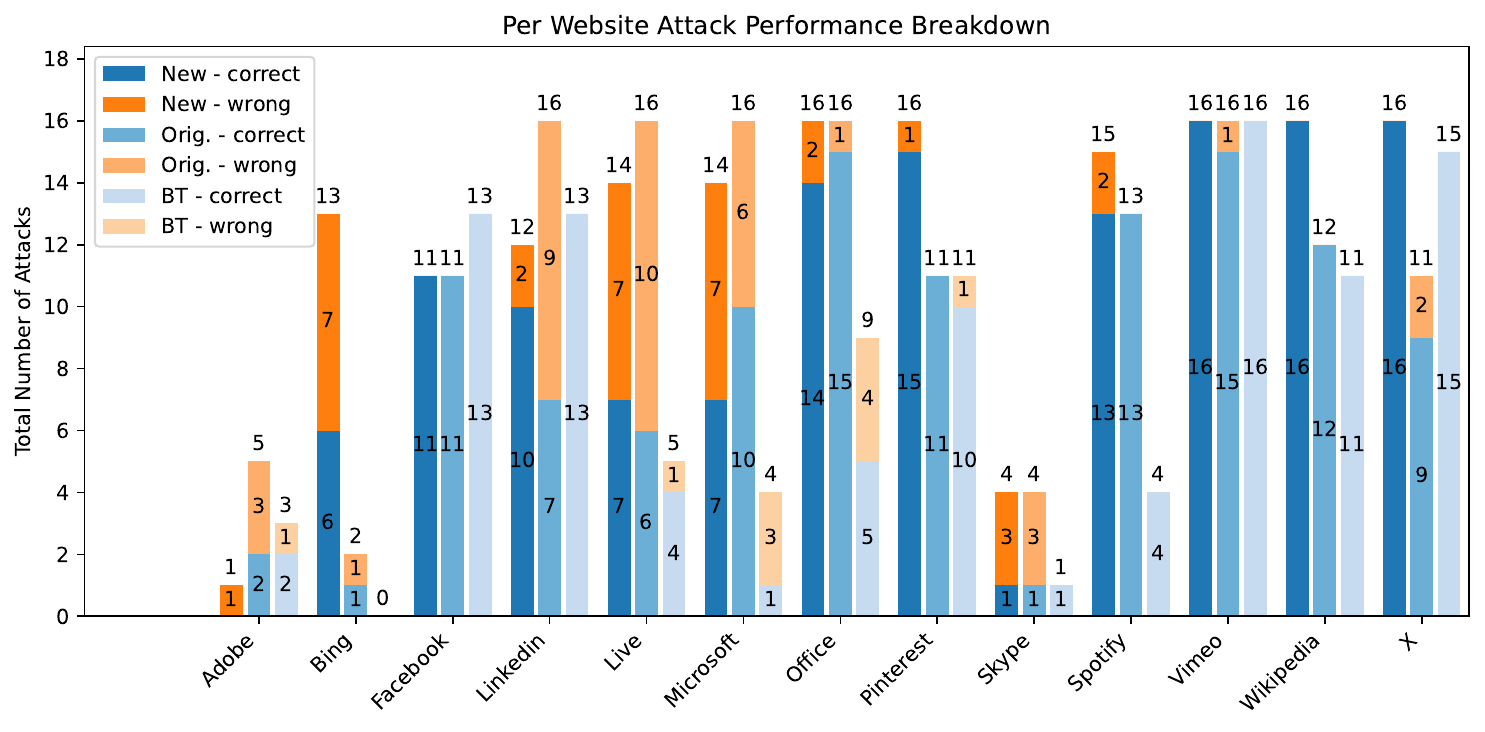}
    \caption{Per website performance of the different attacks (new, original, and BakingTimer).}\label{fig:results_plots}
\end{figure*}

Attacks are conducted on each website at different times of the day to account for fluctuations in the server's load during daily operations. 
Table~\ref{tab:cross-site} reports, aggregated over all tested websites, the number of correct predictions, incorrect predictions, and abstentions produced by each attack. An abstention occurs when the attack fails to predict any class after making the maximum allowed number of requests, due to insufficient statistical evidence.


Overall, the attack based on the new model outperforms both the original model and BakingTimer. Relative to the original model, the new model produces fewer abstentions (-15) and fewer errors (-4), leading to 19 additional correct classifications. In contrast, with respect to BakingTimer, the new model produces a higher number of incorrect classifications, together with fewer abstentions. This disadvantage, however, is only apparent, as BakingTimer's effectiveness is substantially limited by its high number of abstentions, refraining from providing a prediction in 50\% of the cases. Furthermore, among the 59 instances in which the new model provides a classification while BakingTimer abstains, the new model is correct in 37 cases (63\%). This indicates that the additional classifications enabled by the new model are not obtained at the expense of reliability, but instead provide correct predictions in the majority of cases. Indeed, when considering the entire set of 208 attack attempts, the new model yields correct predictions in 63\% of cases, whereas the original model and BakingTimer achieve only 54\% and 46\%, respectively. Importantly, these improvements are obtained with fewer hyper-parameters and less tuning effort, as previously discussed.



\begin{table}[t]
\centering
\caption{Comparison of our attack with the new model, our attack with the original model and BakingTimer.}
\begin{tabular}{lrrr}
\toprule
\textbf{Answer} & \textbf{New Model} & \textbf{Original Model~\cite{ourpets}} & \textbf{BakingTimer~\cite{bakingTimer}} \\
\midrule
Correct            & 132               & 113 & 95                \\
Wrong            & 32                 & 36 & 10               \\
Abstain             & 44                 & 59 & 103               \\
\bottomrule
\end{tabular}
\label{tab:cross-site}
\end{table}


Figure~\ref{fig:results_plots} shows, for each website, the total number of predictions produced by each method, split into correct (blue) and incorrect (orange) outcomes. All three methods perform poorly on Adobe and Skype, suggesting that these sites are not vulnerable to this type of attack, as the response times between the different classes of requests are likely too close to each other. For the remaining websites, the new model either achieves performance comparable to that of the best performing method between the state of the art and the original model, or it substantially increases the number of correct predictions while reducing the number of abstentions. These per-website results are consistent with the aggregate trend observed above: overall, the new model exhibits the lowest number of abstentions or a number comparable to the baseline with the fewest abstentions. Moreover, this reduction in abstentions translates into a higher number of correct predictions rather than incorrect ones. Indeed, across all websites, the new model is either on par with the best performing baseline in terms of correct predictions or it yields a higher number of correct classifications. This further confirms the previously observed trend: the additional classifications enabled by the new model are correct in the majority of cases. The only exception is LinkedIn, where the best performing baseline outperforms the new model, achieving one fewer abstention, two fewer incorrect predictions, and three more correct predictions. Nevertheless, the new model still manages to perform better than the original one.


\section{Related Work}

Remote timing attacks received significant attention from the research community. Felten and Schneider showed that it is possible to abuse the performance improvement offered by web caches to determine whether a target user visited a given web page~\cite{timingAttacksWebPr}. More recent work similarly abused web caches to reconstruct the geo-location of a target user~\cite{JiaDLS15}. Our work does not directly apply to cache-based timing attacks, because the timing side-channel modeled by our framework is the time required for server-side computation. The pioneers of remote timing attacks enabled by the server-side computation time were Bortz and Boneh, who presented different attacks against web applications~\cite{BortzB07}. More recent work generalized their ideas to cover additional attack vectors introduced in modern browsers~\cite{GoethemJN15,0001KK15} and proposed new timing attacks designed to more effectively breach users' privacy~\cite{GelernterH15,bakingTimer}. A key difference between our work and existing attacks is the use of formal modeling to deal with the complexities of remote timing attacks, as discussed in Section~\ref{sec:contribution}. As a matter of fact, our experimental evaluation showed that the effectiveness of empirical state-of-the-art attacks like BakingTimer~\cite{bakingTimer} is sub-optimal. Our attack is more robust to variations in the load factor of the target web application and significantly more precise even when tested against real-world websites.

The only paper taking a more rigorous look into the formal theory of remote timing attacks is the seminal work by Crosby et al.~\cite{CrosbyWR09}.
Here, the authors recognize the importance of accounting for the server load factor when measuring response times, and this impact is evaluated as an additive jitter effect on the measurements. However, the scenario proposed in~\cite{CrosbyWR09} is significantly different from ours. First, the service time of each class is assumed to be a known constant. While this assumption may be reasonable for the type of observations considered in their setting, it is less suitable for web services, where processing times can vary depending on the system state, e.g., due to resource contention and variations in the workload~\cite{gardner}. Second, the jitter is modeled as an additive component of the service time, an assumption that is more appropriate for a First-In First-Out scheduling rather than a Round-robin as we consider here~\cite{harrison}. Round-robin is more accurate in describing the real behavior of schedulers in nowadays web servers~\cite{harchol2000implementation}. 


\section{Conclusion}

In this article, we proposed the use of stochastic models to improve the effectiveness of remote timing attacks and improved upon the work presented in~\cite{ourpets}. Our approach starts from a queuing model of the target web application to attack and uses Bayesian statistics to infer the unknown class of specific network requests of interest. Compared to prior empirical approaches, our attack is more precise and provides quantitative probability bounds. As future work, we would like to experiment with other types of queuing models besides the M/M/1/PS model considered in this article. In particular, we plan to further investigate the challenges posed by microservice architectures.
Moreover, we will explore possibilities to enhance our analytical model by incorporating more complex queuing network frameworks. In fact, supporting different types of models would be important to generalize to different web applications and settings, e.g., the attacker may perform preliminary tests to determine what is the most appropriate model of the target web application. We also plan to improve our experimental evaluation by extending its scale and by testing the robustness of our method against different disruptive factors, e.g., to assess the impact of geo-location on the effectiveness of our attack.

\section*{Acknowledgments}
This work was supported by Agenzia per la cybersicurezza nazionale under the programme for promotion of XL cycle PhD research in cybersecurity – CUP N.H71J24001710005. The views expressed are those of the authors and do not represent the funding institution.

\bibliographystyle{IEEEtran}
\bibliography{IEEEabrv,biblio-toc}

@article{muntz-lstPSwaitTime,
    author = {E. G. Coffman and R. R. Muntz and H. Trotter},
    title = {Waiting Time Distributions for Processor-Sharing Systems},
    year = {1970},
    volume = {17},
    journal = {Journal of ACM},
    publisher = {ACM}
}

@article{morrison,
 author = {J. A. Morrison},
 journal = {SIAM Journal on Applied Mathematics},
 pages = {152--167},
 publisher = {Society for Industrial and Applied Mathematics},
 title = {Response-Time Distribution for a Processor-Sharing System},
 volume = {45},
 year = {1985}
}

@article{RSA78,
 author = {R. Rivest and A. Shamir and L. Adleman},
 title = {A Method for Obtaining Digital Signatures and Public-Key Cryptosystems}, 
 journal = {Comm. of the ACM},
 publisher = {ACM},
 volume = {21},
 pages = {120–126},
 year = {1978}
}

@book{nist,
  title={NIST Handbook of Mathematical Functions Paperback and CD-ROM},
  author={Olver, F.W.J. and National Institute of Standards and Technology (U.S.)},
  year={2010},
  publisher={Cambridge University Press}
}

@article{baccelli2009,
  author       = {Fran{\c{c}}ois Baccelli and Bruno Kauffmann and Darryl Veitch},
  title        = {Inverse problems in queueing theory and Internet probing},
  journal      = {Queueing Syst. Theory Appl.},
  volume       = {63},
  pages        = {59--107},
  year         = {2009},
}

@article{casale18,
  author = {Weikun Wang and Giuliano Casale and  Ajay Kattepur and Manoj Nambiar},
  title = {{QMLE:} {A} Methodology for Statistical Inference of Service Demands from Queueing Data},
  journal = {{ACM} Trans. Model. Perform. Evaluation Comput. Syst.},
  volume       = {3},
  pages        = {17:1--17:28},
  year         = {2018},
}

@article{horvath20,
  author       = {G{\'{a}}bor Horv{\'{a}}th and
                  Ill{\'{e}}s Horv{\'{a}}th and
                  Salah Al{-}Deen Almousa and
                  Mikl{\'{o}}s Telek},
  title        = {Numerical inverse Laplace transformation using concentrated matrix
                  exponential distributions},
  journal      = {Perform. Evaluation},
  volume       = {137},
  year         = {2020},
  pages = {1--22},
}

@article{gardner,
  author       = {Kristen Gardner and
                  Mor Harchol{-}Balter and
                  Alan Scheller{-}Wolf and
                  Mark Velednitsky and
                  Samuel Zbarsky},
  title        = {Redundancy-d: The Power of d Choices for Redundancy},
  journal      = {Oper. Res.},
  volume       = {65},
  pages        = {1078--1094},
  year         = {2017},
}

@article{CrosbyWR09,
  author       = {Scott A. Crosby and
                  Dan S. Wallach and
                  Rudolf H. Riedi},
  title        = {Opportunities and Limits of Remote Timing Attacks},
  journal      = {{ACM} Trans. Inf. Syst. Secur.},
  volume       = {12},
  pages        = {17:1--17:29},
  year         = {2009},
}

@article{BrumleyB05,
  author       = {David Brumley and
                  Dan Boneh},
  title        = {Remote timing attacks are practical},
  journal      = {Comput. Networks},
  volume       = {48},
  pages        = {701--716},
  year         = {2005},
}

@article{JiaDLS15,
  author       = {Yaoqi Jia and
                  Xinshu Dong and
                  Zhenkai Liang and
                  Prateek Saxena},
  title        = {I Know Where You've Been: Geo-Inference Attacks via the Browser Cache},
  journal      = {{IEEE} Internet Comput.},
  volume       = {19},
  pages        = {44--53},
  year         = {2015},
 }

@book{menasce,
   title = {Performance by Design},
   author = {Daniel A. Menasc\'e and Virgilio Almeida and Lawrence W. Dowdy},
   year = {2004},
   publisher = {Pearson}
}

@book{harrison,
author = {Harrison, Peter G. and Patel, Naresh M.},
title = {Performance Modelling of Communication Networks and Computer Architectures (International Computer S},
year = {1992},
isbn = {0201544199},
publisher = {Addison-Wesley Longman Publishing Co., Inc.},
address = {USA},
edition = {1st}
}

@book{kleinrock-vol1,
	author = {Leonard Kleinrock},
	title = {Queueing Systems, Volume 1: Theory},
	year = {1975},
	publisher = {Wiley-Interscience},
}

@book{mor,
	author = {Mor Harchol{-}Balter},
	title = {Performance Modeling and Design of Computer Systems: Queueing Theory in Action},
	year = {2013},
	publisher = {Cambridge University Press},
}

@techreport{harchol2000implementation,
  title={Implementation of SRPT scheduling in web servers},
  author={Harchol-Balter, Mor and Bansal, Nikhil and Schroeder, Bianca and Agrawal, Mukesh},
  year={2000},
  institution={School of Computer Science, Carnegie Mellon University}
}

@inproceedings{cao,
    author = {Jianhua Cao and Mikael Andersson and Christian Nyberg and Maria Kihl},
    year = {2003},
    pages = {1501 - 1506 vol.2},
    title = {Web server performance modeling using an M/G/1/K*PS queue},
    volume = {2},
    booktitle = {Proceedings of the 10th International Conference on Telecommunications (ICT)},
    publisher = {IEEE},
}

@inproceedings{timingAttacksWebPr,
    author       = {Edward W. Felten and
                  Michael A. Schneider},
    title        = {Timing attacks on Web privacy},
    booktitle    = {{CCS} 2000, Proceedings of the 7th {ACM} Conference on Computer and
                  Communications Security, Athens, Greece, November 1-4, 2000},
    pages        = {25--32},
    publisher    = {{ACM}},
    year         = {2000}
}

@inproceedings{BortzB07,
  author       = {Andrew Bortz and
                  Dan Boneh},
  title        = {Exposing private information by timing web applications},
  booktitle    = {Proceedings of the 16th International Conference on World Wide Web,
                  {WWW} 2007, Banff, Alberta, Canada, May 8-12, 2007},
  pages        = {621--628},
  publisher    = {{ACM}},
  year         = {2007},
}

@inproceedings{bakingTimer,
  author       = {Iskander S{\'{a}}nchez{-}Rola and
                  Davide Balzarotti and
                  Igor Santos},
  title        = {BakingTimer: privacy analysis of server-side request processing time},
  booktitle    = {Proceedings of the 35th Annual Computer Security Applications Conference,
                  {ACSAC} 2019, San Juan, PR, USA, December 09-13, 2019},
  pages        = {478--488},
  publisher    = {{ACM}},
  year         = {2019} 
}

@inproceedings{AkhaweBLMS10,
  author       = {Devdatta Akhawe and
                  Adam Barth and
                  Peifung E. Lam and
                  John C. Mitchell and
                  Dawn Song},
  title        = {Towards a Formal Foundation of Web Security},
  booktitle    = {Proceedings of the 23rd {IEEE} Computer Security Foundations Symposium,
                  {CSF} 2010, Edinburgh, United Kingdom, July 17-19, 2010},
  pages        = {290--304},
  publisher    = {{IEEE} Computer Society},
  year         = {2010},
}

@book{Henry11,
  author = {Henry H. Liu},
  title = {Oracle Database Performance and Scalability: A Quantitative Approach},
  year = {2011},
  publisher = {Wiley-IEEE Computer Society Press}
}

@book{ConnieW01,
  author = {Connie U. Smith and Lloyd G. Williams},
  title = {Performance Solutions: A Practical Guide to Creating Responsive, Scalable Software},
  year = {2001},
  publisher = {Addison-Wesley Professional}
}

@inproceedings{PochatGTKJ19,
  author       = {Victor Le Pochat and
                  Tom van Goethem and
                  Samaneh Tajalizadehkhoob and
                  Maciej Korczyński and
                  Wouter Joosen},
  title        = {Tranco: {A} Research-Oriented Top Sites Ranking Hardened Against Manipulation},
  booktitle    = {26th Annual Network and Distributed System Security Symposium, {NDSS}
                  2019, San Diego, California, USA, February 24-27, 2019},
  publisher    = {The Internet Society},
  year         = {2019},
}

@inproceedings{GoethemJN15,
  author       = {Tom van Goethem and
                  Wouter Joosen and
                  Nick Nikiforakis},
  title        = {The Clock is Still Ticking: Timing Attacks in the Modern Web},
  booktitle    = {Proceedings of the 22nd {ACM} {SIGSAC} Conference on Computer and
                  Communications Security, Denver, CO, USA, October 12-16, 2015},
  pages        = {1382--1393},
  publisher    = {{ACM}},
  year         = {2015},
}

@inproceedings{0001KK15,
  author       = {Sangho Lee and
                  Hyungsub Kim and
                  Jong Kim},
  title        = {Identifying Cross-origin Resource Status Using Application Cache},
  booktitle    = {22nd Annual Network and Distributed System Security Symposium, {NDSS}
                  2015, San Diego, California, USA, February 8-11, 2015},
  publisher    = {The Internet Society},
  year         = {2015},
  numpages = {12}
}

@inproceedings{GelernterH15,
  author       = {Nethanel Gelernter and
                  Amir Herzberg},
  title        = {Cross-Site Search Attacks},
  booktitle    = {Proceedings of the 22nd {ACM} {SIGSAC} Conference on Computer and
                  Communications Security, Denver, CO, USA, October 12-16, 2015},
  pages        = {1394--1405},
  publisher    = {{ACM}},
  year         = {2015},
}

@inproceedings{KhodayariP22,
  author       = {Soheil Khodayari and
                  Giancarlo Pellegrino},
  title        = {The State of the SameSite: Studying the Usage, Effectiveness, and
                  Adequacy of SameSite Cookies},
  booktitle    = {43rd {IEEE} Symposium on Security and Privacy, {SP} 2022, San Francisco,
                  CA, USA, May 22-26, 2022},
  pages        = {1590--1607},
  publisher    = {{IEEE}},
  year         = {2022},
}

@article{KistowskiHZKH15,
  author       = {J{\'{o}}akim von Kistowski and
                  Nikolas Herbst and
                  Samuel Kounev and
                  Henning Groenda and
                  Christian Stier and
                  Sebastian Lehrig},
  title        = {Modeling and Extracting Load Intensity Profiles},
  journal      = {{ACM} Trans. Auton. Adapt. Syst.},
  volume       = {11},
  pages        = {23:1--23:28},
  year         = {2017},
}

@misc{Artifacts,
    key = {Artifacts},
    author = {Simone Bozzolan and Diletta Olliaro and Stefano Calzavara and Andrea Marin and Gianfranco Balbo and Matteo Sereno},
    title = {Artifacts},
    year = {2025},
    note = {Repository containing all of the artifacts related to this paper},
    howpublished = {\url{https://github.com/Asterius27/SecPerf-Artifacts/tree/journal_extension}}
}

@misc{matlab_ilt,
  author       = {Gabor Horvath},
  title        = {{A CME-based numerical inverse Laplace transformation method}},
  year         = {2024},
  howpublished = {\url{https://it.mathworks.com/matlabcentral/fileexchange/71511-a-cme-based-numerical-inverse-laplace-transformation-method}},
  note         = {MathWorks}
}

@article{ourpets,
  author = {Simone Bozzolan and Diletta Olliaro and Stefano Calzavara and
                  Andrea Marin and Gianfranco Balbo and Matteo Sereno},
  title        = {Stochastic Models for Remote Timing Attacks},
  journal      = {Proc. Priv. Enhancing Technol.},
  volume       = {2025},
  pages        = {545--559},
  year         = {2025},
}

\begin{IEEEbiography}[{\includegraphics[width=1in,height=1.25in,clip,keepaspectratio]{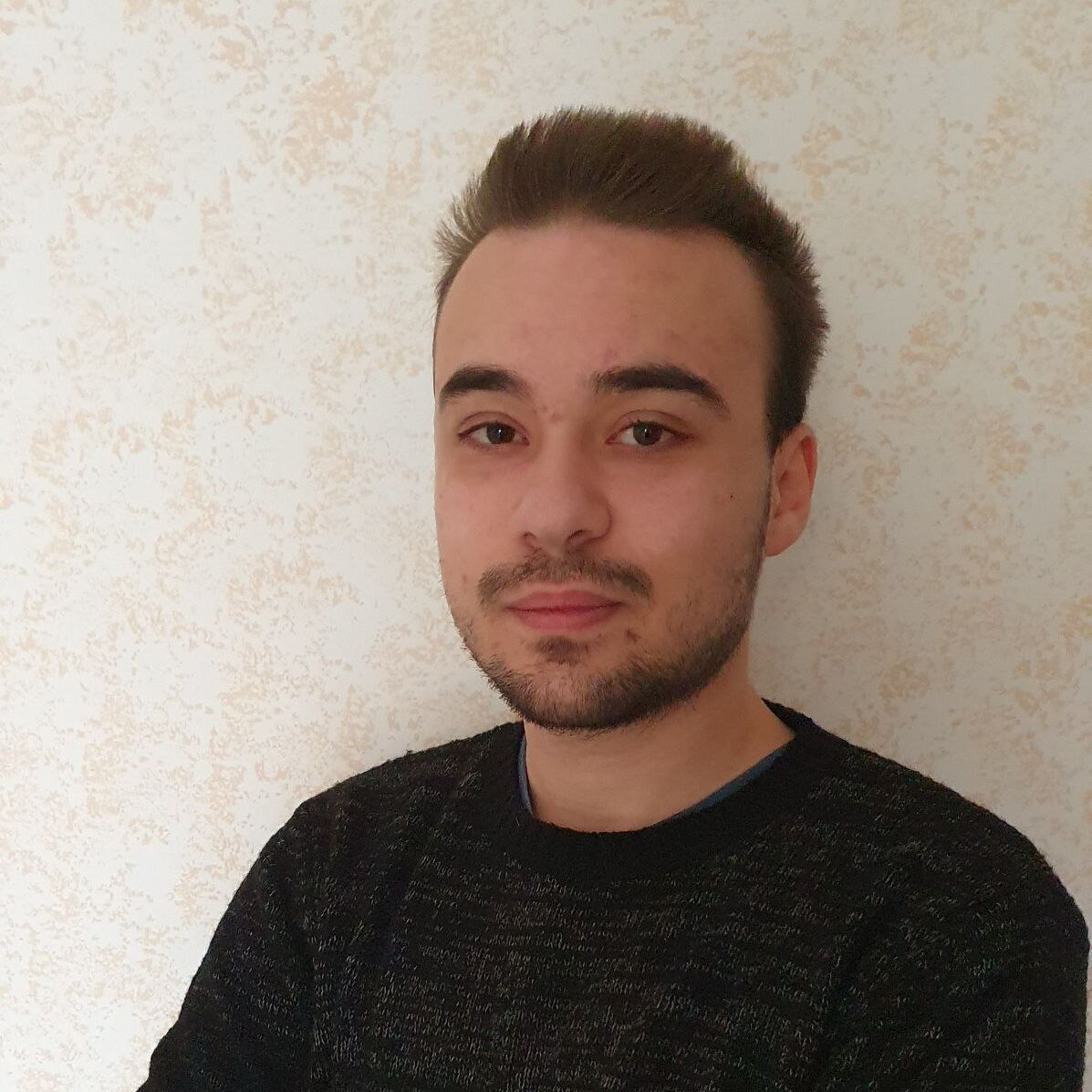}}]{Simone Bozzolan}
received his Master's degree in Computer Science in 2024 from Ca' Foscari University of Venice, where he is currently pursuing a Ph.D. in Computer Science. His research focuses on web security, with particular emphasis on authentication and authorization systems. He investigates techniques for analyzing, detecting, and mitigating vulnerabilities in web applications, applying both empirical and formal methods to improve the security and resilience of web-based systems.
\end{IEEEbiography}
\vspace*{-3\baselineskip}
\begin{IEEEbiography}[{\includegraphics[width=1in,height=1.25in,clip,keepaspectratio]{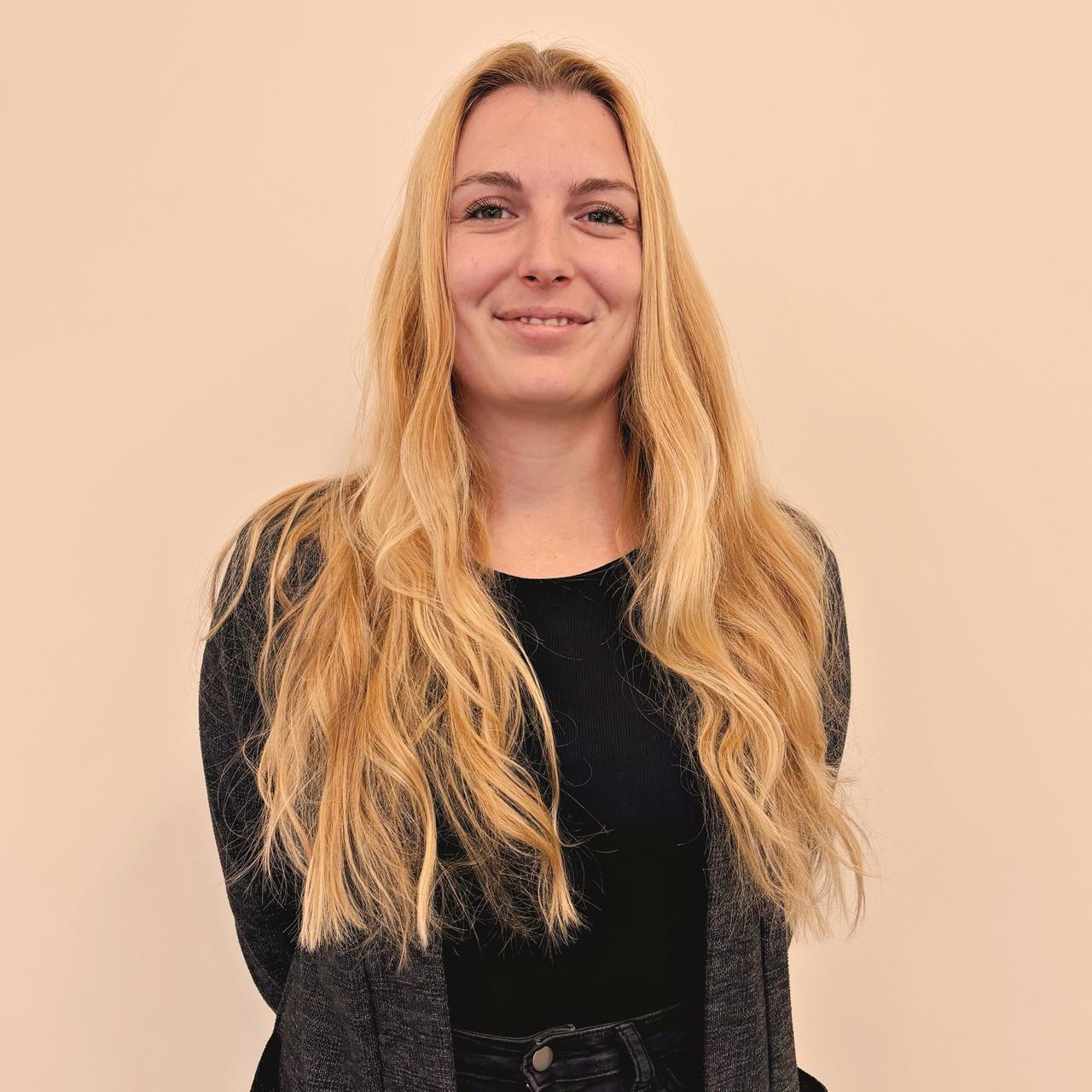}}]{Diletta Olliaro}
is a PostDoctoral Fellow at KTH Royal Institute of Technology where she works in TECoSA research center. In 2025, she was awarded the Ph.D. in Computer Science cum laude from Ca' Foscari University of Venice. Her research focuses on queuing theory, product-form solution models, and, more broadly, the stochastic modeling of networks and telecommunication systems, with an emphasis on evaluating and improving their performance and reliability.
\end{IEEEbiography}
\vspace*{-3\baselineskip}
\begin{IEEEbiography}[{\includegraphics[width=1in,height=1.25in,clip,keepaspectratio]{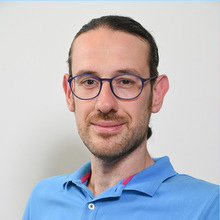}}]{Stefano Calzavara}
is an Associate Professor of Computer Science at Ca’ Foscari University of Venice, Italy. His research lies at the intersection of formal methods and software security, with applications to web security and AI security. He has authored over 70 publications in these areas, including papers in top-tier venues such as IEEE S\&P, ACM CCS, NDSS, USENIX Security, and The Web Conference. He has served the research community in several leading roles, including Program Chair of IEEE CSF (2022 and 2023), Security Area Chair for The Web Conference (2024), and Vice-Chair for USENIX Security (2026 and 2027). He is currently Editor-in-Chief of the Journal of Computer Security. His work has received best paper awards at IEEE S\&P 2025 and ETAPS 2013.
\end{IEEEbiography}
\vspace*{-3\baselineskip}
\begin{IEEEbiography}[{\includegraphics[width=1in,height=1.25in,clip,keepaspectratio]{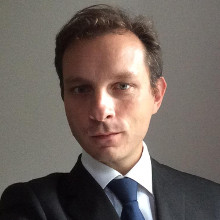}}]{Andrea Marin}
received his Ph.D. in Computer Science from the University of Venice in 2009. He is Professor of Computer Science at the same university. He is the (co-)author of over 100 technical papers in refereed international journals and conference proceedings. His current research focuses on the performance and reliability evaluation of computer systems using stochastic modeling techniques. He is a senior member of IEEE.
\end{IEEEbiography}
\vspace*{-3\baselineskip}
\begin{IEEEbiography}[{\includegraphics[width=1in,height=1.25in,clip,keepaspectratio]{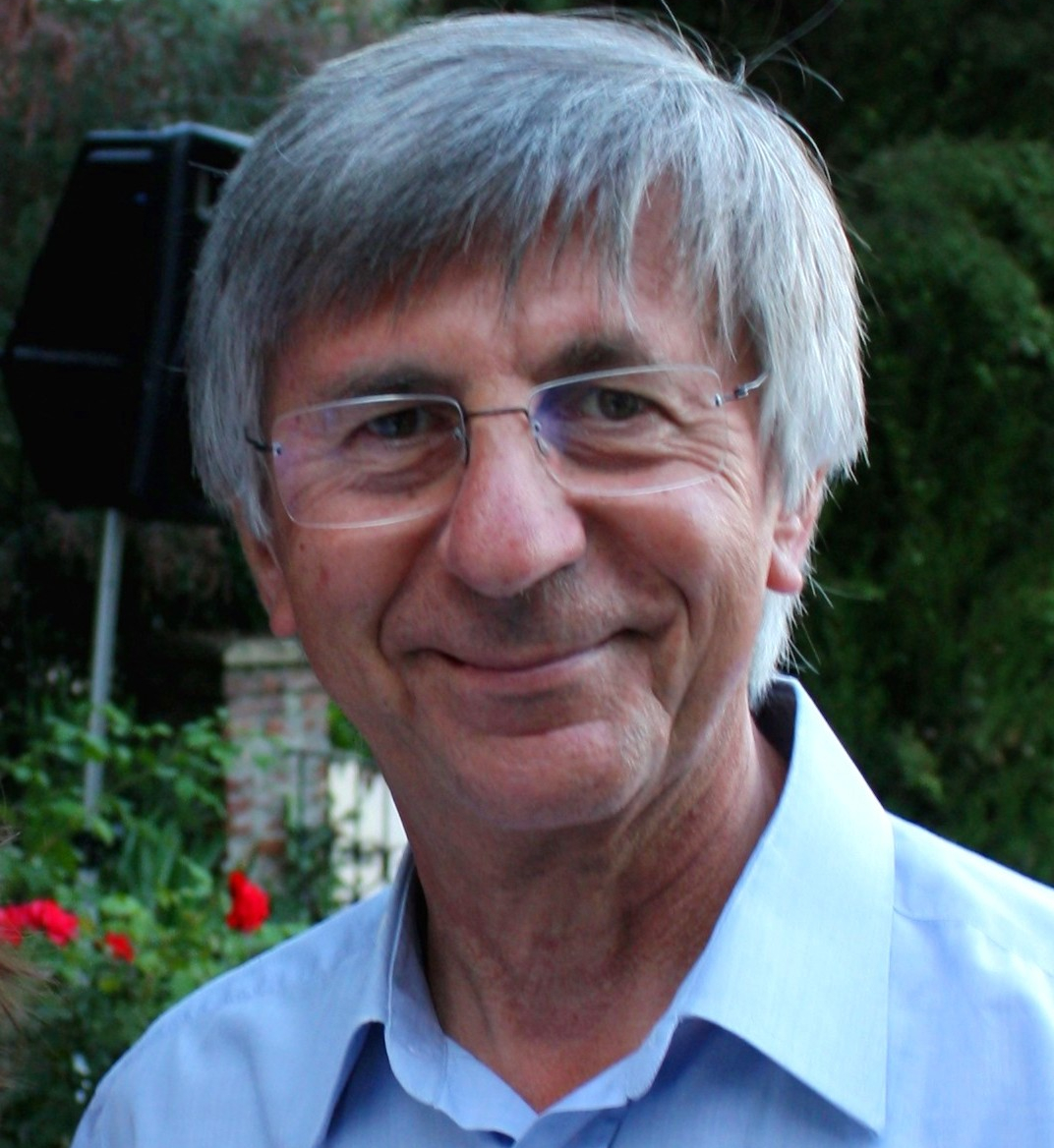}}]{Gianfranco Balbo}
received his Ph.D. degree in Computer Science from Purdue University – USA in 1979 and is currently Emeritus Professor at the Informatics Department of Torino University, after retiring as Full Professor in November 2016. His research interests are in the area of performance evaluation of computer systems, systems biology, queuing networks, stochastic Petri nets, and queuing theory. He is one of the proposers of the Generalized Stochastic Petri Net (GSPN) formalism that he developed while conducting research on the performance of parallel computers. The results of his activity are published in more than 100 conference and journal papers and in 10 books (as author or editor). Gianfranco Balbo is an ISI highly cited researcher and member of the Academy of Sciences of Torino.
\end{IEEEbiography}
\vspace*{-3\baselineskip}
\begin{IEEEbiography}[{\includegraphics[width=1in,height=1.25in,clip,keepaspectratio]{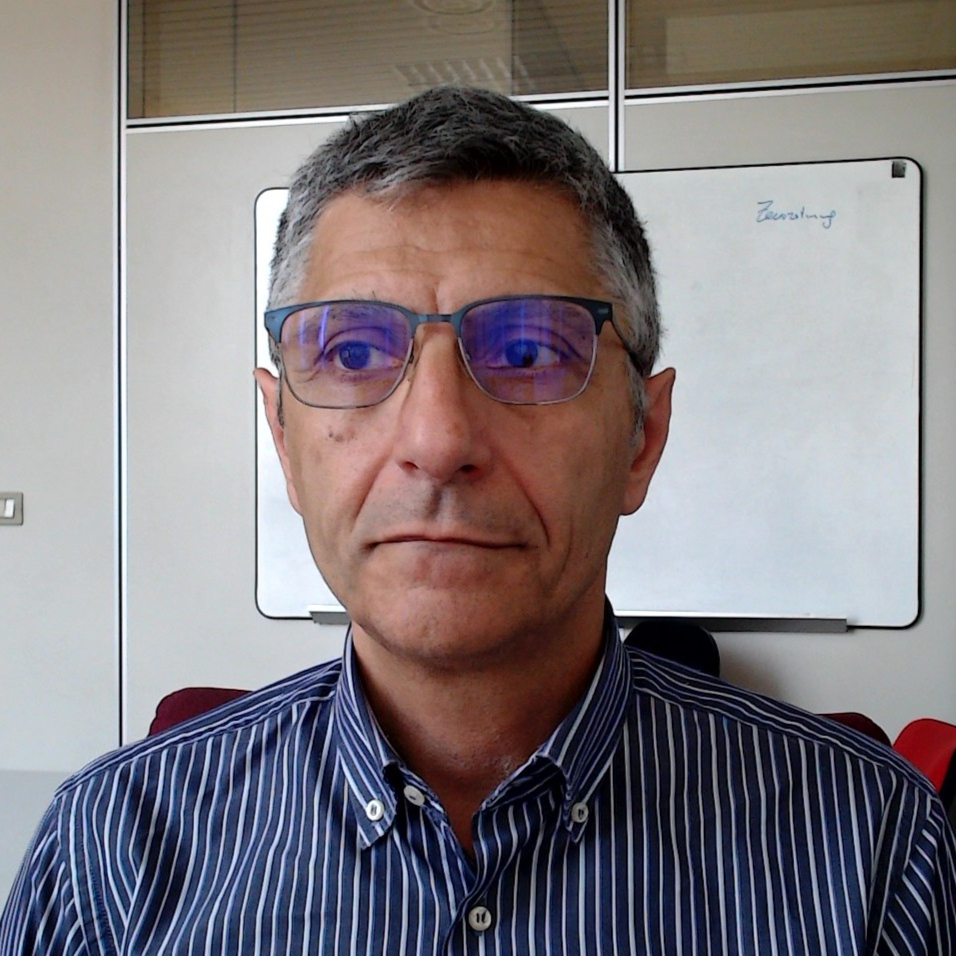}}]{Matteo Sereno}
received the Laurea degree in Computer Science from the University of Salerno, Italy, in 1987, and the Ph.D. degree in Computer Science from the University of Turin, Italy, in 1992. He is currently a Full Professor with the Department of Computer Science, University of Turin, Italy. His research interests include performance evaluation and modeling of communication networks, distributed systems, queuing networks, stochastic Petri net models, game theory, and their applications to computer and telecommunication systems.
\end{IEEEbiography}

\end{document}


\maketitle

\section*{Supplementary Material}

\subsection{Detailed Proof of and Supporting Lemmas for Theorem 1}

\begin{lemma}\label{lem:geomseries}
For a fixed $s \ge 0$ and any $\tau \in \mathbb{R}^+$, the following expansion holds:
\[
    \dfrac{1}{1-\dfrac{g(s)}{b(s)}\exp(-h(s)\tau)} = \sum_{n=0}^{\infty} \left(\dfrac{g(s)}{b(s)}\exp(-h(s)\tau)\right)^n\,.
\]
\end{lemma}
\begin{proof}
The geometric series 
\[
    \sum_{n=0}^{\infty} x^n = \dfrac{1}{1-x}
\]
if and only if $|x| < 1$. Let 
\[
x = \dfrac{g(s)}{b(s)}\exp(-h(s)\tau)\,.
\]
Since $h(s) > 0$ for $s \ge 0$, the exponential term satisfies $0 < \exp(-h(s)\tau) < 1$ for all $\tau > 0$. So, we need $\left|\dfrac{g(s)}{b(s)}\right| < 1$
\begin{align*}
    \dfrac{\rho(1-\muntzroot(s))^2}{(1-\rho \muntzroot(s))^2} &< 1\\
    \rho(1-\muntzroot(s))^2 &< (1-\rho \muntzroot(s))^2\\
    \rho(1-2\muntzroot(s)+\muntzroot(s)^2) &< (1-2\rho \muntzroot(s) + \rho^2 \muntzroot(s)^2) \\
    \rho +\rho \muntzroot(s)^2 -1 -\rho^2 \muntzroot(s)^2 &< 0\\
    \rho -1 + \rho \muntzroot(s)^2 (1-\rho) &< 0\\
    (\rho -1) (1-\rho \muntzroot(s)^2) &< 0
\end{align*}
    since $0<\rho<1$, we have $(\rho -1)<0$, thus the inequality holds if 
    \[
    1-\rho \muntzroot(s)^2 > 0,\quad  \text{i.e.,}\quad \muntzroot(s)^2 <\dfrac{1}{\rho}
    \]
    as $0<\muntzroot(s)\leq 1$ we need $1\leq \rho^{-1}$ which is always true for $0< \rho <1$. Thus, the condition for the convergence of the geometric series is satisfied.
\end{proof}

\begin{lemma}\label{lem:changeOfVar}
For $k, p \in \mathbb{R}^+$, the following integral identity holds:
\[
\int_0^{\infty} t^{k-1} \exp(-p t)\ \d t = \dfrac{\Gamma(k)}{p^k}\,.
\]    
\end{lemma}
\begin{proof}
Applying the change of variable $u = pt$, so that $\d t = \dfrac{1}{p}\d u$, we have:
\begin{align*}
    \int_0^{\infty} t^{k-1} \exp(-p t)\ \d t &= \int_0^{\infty} \left(\dfrac{u}{p}\right)^{k-1} \dfrac{\exp(-u)}{p} \ \d u && \text{by substitution } u = p t \\
    &= \dfrac{1}{p^k} \int_0^{\infty} u^{k-1} \exp(-u)\ \d u && \text{factoring out constants} \\
    &= \dfrac{\Gamma(k)}{p^k} && \text{by integral defintion of the Gamma function.}
\end{align*}
\end{proof}

\begin{proof}[Proof of Theorem 1]
    Let us start from Eq. (6), considering $\vcdf(\tau)$ to be the CDF of a Gamma distribution. Differentiating the function inside the measure for the Lebesgue integral we obtain: 
    \[
    \vcdf(\tau)=\dfrac{(k\gamma)^k(\tau^{k-1}\exp(-k \gamma \tau))}{(k-1)!}\,, \quad \text{with } \gamma \in \mathbb{R}^+\,.
    \]
    
    Now, we substitute $\rlst(\tau, s)$ with Eq. (5) in Eq. (6) and retrieve:
   
\begin{equation}\label{eq:rs_riemann_erl}
    \rlst(s) = \int_0^{\infty} \exp(-s\tau)\wlst(\tau, s) \dfrac{(k\gamma)^k(\tau^{k-1}\exp(-k \gamma \tau))}{(k-1)!}\ d\tau
\end{equation}
From which we obtain:
\begin{align*}
    \rlst(s) &= \dfrac{(k\gamma)^k (1-\rho)(1-\rho \muntzroot(s)^2)}{(k-1)!} \int_0^{\infty} \exp(-s\tau) \dfrac{\exp(-a(s)\tau)}{b(s)-g(s)\cdot \exp(-h(s)\tau)} \tau^{k-1}\exp(-k \gamma \tau)\ \d \tau\\
    &= \dfrac{(k\gamma)^k (1-\rho)(1-\rho \muntzroot(s)^2)}{(k-1)!} \int_0^{\infty}  \dfrac{\tau^{k-1} \exp(-(s+a(s)+k\gamma)\tau)}{b(s)-g(s)\cdot \exp(-h(s)\tau)} \ \d\tau \\
    &= \dfrac{(k\gamma)^k (1-\rho)(1-\rho \muntzroot(s)^2)}{(k-1)!} \int_0^{\infty} \tau^{k-1} \exp(-(s+a(s)+k\gamma)\tau) \left[\dfrac{1}{b(s)}\sum_{n=0}^{\infty}\left( \dfrac{g(s)}{b(s)} \exp(-h(s)\tau) \right)^n \right] \ \d\tau\quad \text{by Lemma}~\ref{lem:geomseries}\\\\
    &= \dfrac{(k\gamma)^k (1-\rho)(1-\rho \muntzroot(s)^2)}{(k-1)!} \dfrac{1}{b(s)}\sum_{n=0}^{\infty} \dfrac{g(s)^n}{b(s)^{n}} \int_0^{\infty} \tau^{k-1} \exp(-(s+a(s)+k\gamma+nh(s))\tau)  \ \d\tau\\
    &= \dfrac{(k\gamma)^k (1-\rho)(1-\rho \muntzroot(s)^2)}{(k-1)!} \dfrac{1}{b(s)}\sum_{n=0}^{\infty} \dfrac{g(s)^n}{b(s)^{n}} \dfrac{\Gamma(k)}{(s+a(s)+k\gamma + nh(s))^k}\quad \text{by Lemma}~\ref{lem:changeOfVar}\\
    &= \dfrac{(k\gamma)^k (1-\rho)(1-\rho \muntzroot(s)^2)}{\Gamma(k)} \dfrac{1}{b(s)(s)}\sum_{n=0}^{\infty} \dfrac{g(s)^n}{b(s)^{n}} \dfrac{\Gamma(k)}{(s+a(s)+k\gamma + nh(s))^k}\quad \text{by Definition of Gamma function}\\
    &= (k\gamma)^k (1-\rho)(1-\rho \muntzroot^2) \dfrac{1}{b}\sum_{n=0}^{\infty} \dfrac{g(s)^n}{b(s)^{n}} \dfrac{1}{(s+a(s)+k\gamma + nh(s))^k}\quad \text{simplifying Gamma function}\\
    &= (k\gamma)^k (1-\rho)(1-\rho \muntzroot(s)^2) \dfrac{1}{b(s)h(s)^k}\Phi\left(\dfrac{g(s)}{b(s)}, k, \dfrac{s+a(s)+k\gamma}{h(s)}\right)\quad \text{by }~\cite[\text{Eq.}~(25.14.1), \text{pp.}~612]{nist}
\end{align*}
\end{proof}


\subsection{Algebraic Derivation of Eq. (8)}
To compute the variance of the response time $\mathcal{R}$, we evaluate:
\begin{align*}
     \text{Var}({\mathcal R}) &=\mathbb{E}[{\mathcal R}^2] - (\mathbb{E}[{\mathcal R}])^2 \\
     &= \left.\dfrac{\d^2}{\d s^2}\, \rlst(s) \right|_{s=0} - \left[\left.\dfrac{\d}{\d s}\, \rlst(s) \right|_{s=0}\right]^2
\end{align*}
The first and second derivatives of the individual components comprising the response time transform are obtained via standard algebraic and chain-rule derivations. Their values and derivatives evaluated at the boundary $s=0$ are compactly summarized in Table~\ref{tab:all_derivatives}.

\begin{table}[ht]
    \centering
    \small
    \begin{tabular}{llll}
    \hline
    \textbf{Function} & $f(0)$ & $f'(0)$ & $f''(0)$ \\ \hline
    $y(s)$ & $1$ & $-\dfrac{1}{\mu(1-\rho)}$ & $\dfrac{2}{\mu^2(1-\rho)^3}$ \\ [1ex]
    $a(s)$ & $0$ & $\dfrac{\rho}{1-\rho}$ & $\dfrac{-2\rho}{\mu(1-\rho)^3}$ \\ [1ex]
    $b(s)$ & $(1-\rho)^2$ & $\dfrac{2\rho}{\mu}$ & $\dfrac{2\rho(\rho-2)}{\mu^2(1-\rho)^2}$ \\ [1ex]
    $g(s)$ & $0$ & $0$ & $\dfrac{2\rho}{\mu^2(1-\rho)^2}$ \\ [1ex]
    $h(s)$ & $\mu(1-\rho)$ & $\dfrac{1+\rho}{1-\rho}$ & $\dfrac{-4\rho}{\mu(1-\rho)^3}$ \\ [1ex]
    $z(s)$ & $0$ & $0$ & $\dfrac{2\rho}{\mu^2(1-\rho)^4}$ \\ [1ex]
    $M(s)$ & $k\gamma$ & $\dfrac{1}{1-\rho}$ & $\dfrac{-2\rho}{\mu(1-\rho)^3}$ \\ [1ex]
    $\alpha(s)$ & $\dfrac{k\gamma}{\mu(1-\rho)}$ & $\dfrac{\mu(1-\rho)-k\gamma(1+\rho)}{\mu^2(1-\rho)^3}$ & $\dfrac{-2\mu(1+2\rho)(1-\rho) + 2k\gamma(\rho^2+4\rho+1)}{\mu^3(1-\rho)^5}$ \\ \hline
    \end{tabular}
    \caption{Summary of foundational functions and their derivatives evaluated at $s=0$.}
    \label{tab:all_derivatives}
\end{table}

\subsubsection{Derivatives of the Lerch Transcendent Components}
Let $z(s) = \dfrac{g(s)}{b(s)}$, $\alpha(s) = \dfrac{s+a(s)+k\gamma}{h(s)}$, and $\Phi_k(s) = \Phi(z(s), k, \alpha(s))$. At $s=0$, we have $\Phi_k(0) = \left(\dfrac{\mu(1-\rho)}{k\gamma}\right)^k$. 
To compute the derivatives of $\Phi_k(s)$, we apply the multivariate chain rule using the standard partial derivatives of the Lerch Transcendent. Due to an indeterminate form when evaluating $\dfrac{\partial \Phi}{\partial z}$ at $s=0$, we leverage its infinite series definition:
\[
\left.\dfrac{\partial \Phi_k(s)}{\partial z(s)}\right|_{s=0} = \sum_{n=1}^\infty \dfrac{nz(0)^{n-1}}{(n+\alpha(0))^k} = \dfrac{1}{(1+\alpha(0))^k} = \left[\dfrac{\mu(1-\rho)}{\mu(1-\rho)+k\gamma}\right]^k
\]
The remaining first and second partial derivatives with respect to $\alpha(s)$ at $s=0$ yield:
\[
\left.\dfrac{\partial \Phi_k(s)}{\partial \alpha(s)}\right|_{s=0} = -k\left[\dfrac{\mu(1-\rho)}{k\gamma}\right]^{k+1}, \quad \text{and} \quad \left.\dfrac{\partial^2 \Phi_k(s)}{\partial (\alpha(s))^2}\right|_{s=0} = k(k+1)\left[\dfrac{\mu(1-\rho)}{k\gamma}\right]^{k+2}
\]
Applying the total derivative with respect to $s$ at $s=0$, we find:
\begin{align*}
    \dfrac{\d}{\d s} \Phi_k(0) &= -k \alpha^{'}(0) \Phi_{k+1}(0) = \dfrac{-k(\mu(1-\rho)-k\gamma(1+\rho))(\mu^{k-1}(1-\rho)^{k-2})}{(k\gamma)^{k+1}} \\
    \dfrac{\d^2}{\d s^2} \Phi_k(0) &= \dfrac{\partial^2\Phi_k(0)}{\partial(\alpha(0))^2}(\alpha^{'}(0))^2 + \dfrac{\partial \Phi_k(0)}{\partial z(0)} z^{''}(0) + \dfrac{\partial\Phi_k(0)}{\partial \alpha(0)} \alpha^{''}(0)
\end{align*}

\subsubsection{First and Second Derivatives of the Response Time Transform}
We define the scaling function $W(s) = \dfrac{1-\rho \muntzroot^2}{b(s)h(s)^k}$, which allows us to express the transform compactly as $\mathcal{D}\left[F^*_{\mathcal R}(s)\right] = (k\gamma)^k(1-\rho) W(s)\Phi_k(s)$. Evaluating $W(s)$ and its derivatives at $s=0$ via the quotient rule yields:
\[
W(0) = \dfrac{1}{\mu^k(1-\rho)^{k+1}}, \quad W'(0) = \dfrac{-k(1+\rho)}{\mu^{k+1}(1-\rho)^{k+3}}, \quad W''(0) = \dfrac{-2\rho+4k\rho +k(k+1)(1+\rho)^2}{\mu^{k+2}(1-\rho)^{k+5}}
\]

\paragraph{First Derivative Evaluation:}
Using the product rule, the first derivative at $s=0$ simplifies beautifully:
\begin{align*}
    \mathcal{D}^{'}\left[F^*_{\mathcal R}(0)\right] &= (k\gamma)^k(1-\rho) \left[W^{'}(0) \Phi_k(0) + W(0)\dfrac{\d}{\d s}\Phi_k(0)\right] = -\dfrac{1}{\gamma(1-\rho)}
\end{align*}
Thus, the first moment is $-\mathcal{D}^{'}\left[F^*_{\mathcal R}(0)\right] = \dfrac{1}{\gamma(1-\rho)}$.

\paragraph{Second Derivative Evaluation:}
Differentiating a second time yields the second moment component:
\begin{align*}
    \mathcal{D}^{''}\left[F^*_{\mathcal R}(0)\right] &= (k\gamma)^k(1-\rho) \left[W^{''}(0) \Phi_k(0)+ 2W^{'}(0)\dfrac{\d}{\d s}\Phi_k(0) + W(0)\dfrac{\d^2}{\d s^2} \Phi_k(0)\right] \\
    &= \dfrac{1}{\mu^2(1-\rho)^4} \Bigg[ -2\rho + \dfrac{2\rho\mu(1-\rho)}{\gamma} + \dfrac{(k+1)\mu^2(1-\rho)^2}{k\gamma^2} + 2\rho\left[\dfrac{k\gamma}{\mu(1-\rho)+k\gamma}\right]^k \Bigg]
\end{align*}

\subsubsection{Final Variance Form}
Substituting the first and second moments back into the variance equation, the mutual terms cancel out elegantly:
\begin{align*}
    \text{Var}({\mathcal R}) &= \mathcal{D}^{''}\left[F^*_{\mathcal R}(0)\right] - \left(-\mathcal{D}^{'}\left[F^*_{\mathcal R}(0)\right]\right)^2 \\
    &= \dfrac{1}{k\gamma^2(1-\rho)^2}+\dfrac{2\rho}{\mu^2(1-\rho)^4} \Bigg[ \dfrac{\mu(1-\rho)}{\gamma} + \left[\dfrac{k\gamma}{\mu(1-\rho)+k\gamma} \right]^k-1 \Bigg]
\end{align*}
Finally, substituting $\rho = \dfrac{\lambda}{\mu}$ and $(1-\rho) = \dfrac{\mu-\lambda}{\mu}$ yields the final closed-form variance expression:
\[
\text{Var}({\mathcal R}) = \dfrac{\mu\left(\mu(\mu-\lambda)^2+2\gamma k\lambda(\mu-\lambda) + 2\gamma^2k\lambda\left(\dfrac{k\gamma}{\mu-\lambda+k\gamma}\right)^k-2\gamma^2 k\lambda\right)}{\gamma^2k(\mu-\lambda)^4}
\]

\bibliographystyle{IEEEtran}
\bibliography{IEEEabrv,biblio-toc}